\documentclass[pre,aps,twocolumn]{revtex4}
\usepackage{epsfig,latexsym}
\usepackage{color}
\usepackage[hypertex]{hyperref}
\usepackage{bm}

\begin{document}

\preprint{APS/123-QED}

\title{Critical dynamics of self-gravitating Langevin particles and
bacterial populations}

\author{Cl\'ement Sire and Pierre-Henri Chavanis}
\affiliation{Laboratoire de Physique Th\'eorique - IRSAMC, CNRS, Universit\'e Paul Sabatier,
31062 Toulouse, France}

\date{\today}

\begin{abstract}
\vskip0.3cm
We study the critical dynamics of the generalized Smoluchowski-Poisson
system (for self-gravitating Langevin particles) or generalized
Keller-Segel model (for the chemotaxis of bacterial populations).
These models [Chavanis \& Sire, PRE, {\bf 69}, 016116 (2004)] are
based on generalized stochastic processes leading to the Tsallis
statistics. The equilibrium states correspond to polytropic
configurations with index $n$ similar to polytropic stars in
astrophysics. At the critical index $n_{3}=d/(d-2)$ (where $d\ge 2$ is
the dimension of space), there exists a critical temperature
$\Theta_{c}$ (for a given mass) or a critical mass $M_{c}$ (for a
given temperature). For $\Theta>\Theta_{c}$ or $M<M_{c}$ the system
tends to an incomplete polytrope confined by the box (in a bounded
domain) or evaporates (in an unbounded domain).  For
$\Theta<\Theta_{c}$ or $M>M_{c}$ the system collapses and forms, in a
finite time, a Dirac peak containing a finite fraction $M_c$ of the
total mass surrounded by a halo. We study these regimes numerically
and, when possible, analytically by looking for self-similar or pseudo
self-similar solutions. This study extends the critical dynamics of
the ordinary Smoluchowski-Poisson system and Keller-Segel model in
$d=2$ corresponding to isothermal configurations with
$n_{3}\rightarrow +\infty$.  We also stress the analogy between the
limiting mass of white dwarf stars (Chandrasekhar's limit) and the
critical mass of bacterial populations in the generalized Keller-Segel
model of chemotaxis.
\end{abstract}

\pacs{???}
\maketitle

\section{Introduction}
\label{sec_introduction}

For a long time, statistical mechanics was restricted to systems
interacting via short-range forces. For example, the case of
self-gravitating systems is almost never considered in standard
textbooks of statistical mechanics and these systems have been studied
exclusively in the context of astrophysics. In the sixties, Antonov
\cite{antonov}, Lynden-Bell \cite{lb} and Thirring \cite{thirring}
realized that self-gravitating systems have a very special
thermodynamics marked by the non-equivalence of statistical ensembles
(microcanonical, canonical, grand canonical,...). This is related to
the non-additivity of the energy and to the presence of negative
specific heats in the microcanonical ensemble. Furthermore, these
systems experience a rich diversity of phase transitions
(microcanonical and canonical first order phase transitions, zeroth
order phase transitions,...) associated with their natural tendency to
undergo {gravitational collapse}
\cite{paddy,ijmpb}. Recently, several researchers have started to
consider the dynamics and thermodynamics of systems with long-range
interactions at a more general level (see the books \cite{houches,assise} and
references therein) and to discuss the numerous analogies (and
differences) between these systems: self-gravitating systems,
two-dimensional vortices, neutral and non-neutral plasmas, the HMF
model, free electron lasers, Bose-Einstein condensates, atomic
clusters, chemotaxis of bacterial populations etc. These analogies
have also suggested interesting experiments. For example, in the
physics of ultra-cold gases, some authors \cite{artemiev} have
proposed to generate an attractive $1/r$ interaction between atoms by
using a clever configuration of laser beams. This leads to the
fascinating possibility of reproducing, in the laboratory, the {\it
isothermal collapse} (in the canonical ensemble) of a self-gravitating
Fermi gas \cite{ht,pt} leading to a ``white dwarf star''. These
examples illustrate the importance of studying the statistical
mechanics of systems with long-range interactions at a general level
and to develop the analogies between different systems that may seem
{\it a priori} of a very different nature.

\begin{table*}[ht]
\begin{tabular}
{||c||c|c|c||c||}%
\hline\hline
   Index & Temperature  &    Bounded domain      &  Unbounded domain   \\
 \hline \hline
     & $T>T_c$     & Metastable equilibrium state & $\bullet$ Evaporation \cite{virial1}: \\
     &     &  (local minimum of free energy): & asymptotically free normal   \\
 $n=\infty$  &   & box-confined isothermal sphere \cite{brenner,crs,sc}  & diffusion (gravity negligible)\\
  \cline{2-3}
     & $T<T_c$     & Self-similar collapse with $\alpha=2$ \cite{brenner,crs,sc} & $\bullet$ Collapse:   \\
   &     & followed by a self-similar post-collapse leading & pre-collapse  and post-collapse as \\
  &  &  to the  formation of a Dirac peak of mass $M$ \cite{post} & in a bounded domain \cite{brenner,crs,sc} \\
\hline \hline
    & $\Theta>\Theta_c$     & Equilibrium state:  & Equilibrium state:  \\
 $0<n<n_3$     &     & box-confined (incomplete) polytrope \cite{lang} & complete polytrope  \\
\cline{2-3}
      & $\Theta<\Theta_c$     & Equilibrium state:  & (compact support) \cite{lang}  \\
   &     & complete polytrope (compact support) \cite{lang}&    \\
\hline \hline
    & $\Theta>\Theta_c$     & Metastable equilibrium state & $\bullet$ Evaporation [P]:  \\ 
     &     &  (local minimum of free energy): & asymptotically free anomalous   \\
 $n_3<n<\infty$     &     &  box-confined polytropic sphere \cite{lang}&  diffusion (gravity negligible)    \\
\cline{2-3}
      & $\Theta<\Theta_c$     & Self-similar collapse with $\alpha=2n/(n-1)$ \cite{lang}& $\bullet$ Collapse:   \\
   &     & followed by a post-collapse leading to the & pre-collapse and  post-collapse \\
   &     &   formation of a Dirac peak of mass $M$ [N] & as in a bounded domain \cite{lang} \\
\hline \hline
 & $\Theta>\Theta_c$     & Equilibrium state: & Self-similar evaporation \\
 $n=n_3$     &    &  box-confined (incomplete) polytrope \cite{lang} & modified by self-gravity [P]  \\
  \cline{2-4}
     & $\Theta<\Theta_c$     & Pseudo self-similar collapse & Collapse [N]\\
   &     & leading to a Dirac peak of &  \\
   &     & mass $(\Theta/\Theta_c)^{d/2}M$ $+$ halo [P]. &  \\
   &     & This is followed by a post-collapse &  \\
   &     & leading to a Dirac peak of mass $M$ [N] &  \\
\cline{2-4}
     & $\Theta=\Theta_c$     & Infinite family of steady states [P] & Infinite family of steady states [P] \\
\hline \hline
\end{tabular}
\vspace{.3cm}
{\caption{Summary of the different regimes of the GSP system in $d>2$ with references to the physical literature ([P]: present paper; [N]: not done). The case of negative indices is considered in \cite{logotropes}. The links to the mathematical literature are indicated in the main text. {\it Note:} for $(n=\infty, T>T_{c})$ and for $(n_{3}<n<\infty, \Theta>\Theta_{c})$ in a bounded domain, the system can either reach  a metastable equilibrium state or collapse depending on a notion of basin of attraction (see \cite{crs} for more details).} \label{Table1}}
\end{table*}

\begin{table*}[ht]
\begin{tabular}
{||c||c|c|c||c||}%
\hline\hline
   Index & Temperature  &    Bounded domain      &  Unbounded domain   \\
 \hline \hline
 & $T>T_c$     & Equilibrium state: & Self-similar evaporation \\
 $n=\infty$     &    & analytical solution \cite{sc} & modified by self-gravity \cite{virial1}   \\
  \cline{2-4}
     & $T<T_c$     & Pseudo self-similar collapse & Collapse [N] \\
   &     & leading to a Dirac peak of  &  \\
   &     &  mass $(T/T_c)M$     $+$ halo \cite{herrerobio,sc}.                                              &  \\
   &     &  This is followed by a post-collapse             &  \\
   &     &  leading to  a Dirac peak of mass $M$ [N]           &  \\
  \cline{2-4}
     & $T=T_c$     & Self-similar collapse leading to& Self-similar collapse leading to \\
    &     &  a Dirac peak of mass $M$ with &  a Dirac peak of mass $M$ with \\
 &  & exponential growth of $\rho(0,t)$ \cite{sc,souplet} & logarithmic growth of $\rho(0,t)$ \cite{virial1} \\
\hline \hline
    & $T>T_c$     & Equilibrium state:  & Equilibrium state: \\
 $0<n<\infty$     &     & box-confined (incomplete) polytrope \cite{lang} & complete polytrope (compact support)   \\
\cline{2-3}
      & $T\le T_c$     & Equilibrium state:  & \cite{lang}  \\    
   &     & complete polytrope (compact support) \cite{lang}&  \\
\hline \hline
\end{tabular}
\vspace{.3cm}
{\caption{Summary of the different regimes of the GSP system in
$d=2$. In $d=1$, the GSP system always relaxes towards a statistical
equilibrium state so that there is no evaporation or collapse
\cite{sc,acedo,mt}.} \label{Table2}}
\end{table*}

In a series of papers \cite{crs,sc,post,tcoll,multi,virial1,virial2},
we have investigated the dynamics and thermodynamics of a system of
self-gravitating random walkers. The basic idea is to couple the usual
Brownian motion (as introduced by Einstein and Smoluchowski) to the
gravitational interaction.  In our general model \cite{virial2}, the
microscopic dynamics of the particles is described by $N$ coupled
stochastic equations including a friction force and a stochastic force
in addition to the gravitational interaction. The friction force and
the stochastic force model the interaction of the system with a thermal bath of
non-gravitational origin. Then, the proper statistical description of
this {\it dissipative} system is the canonical ensemble. In order to
simplify the problem, we have considered a strong friction limit in
which the motion of the particles is overdamped.  We have also
considered a mean field approximation which becomes exact in a proper
thermodynamic limit $N\rightarrow +\infty$ in such a way that the
volume $V\sim 1$ is of order unity and the coupling constant $G\sim
1/N$ goes to zero (alternatively, we can consider that the mass of the
individual particles scales like $m\sim 1/N$ so that the total mass
$M\sim Nm$ and the gravity constant $G$ remain of order unity). These
approximations lead to the Smoluchowski-Poisson (SP) system. The
steady states correspond to isothermal distributions associated with
the Boltzmann statistics. When coupled to the Poisson equation, we
obtain density profiles similar to {\it isothermal stars} in astrophysics
\cite{emden,chandrab}. In the course of our study, we realized that the SP system is
isomorphic to the standard Keller-Segel (KS) model \cite{ks,jl}
introduced in mathematical biology to describe the chemotaxis of
bacterial populations \cite{murray}. The SP system and the KS model
have now been extensively studied by physicists
\cite{crs,sc,post,tcoll,multi,virial1,virial2,acedo} and applied
mathematicians \cite{nanjundiah,cp,childress,nagai,biler95,herrero96,herrerobio,othmer,herrero97,herrero98,biler,brenner,nagai2,rosier,bn,horstmann,dolbeault,biler4,corrias,biler1,biler2,blanchet1,blanchet2,souplet} with different methods and motivations.

We have also studied a generalized Smoluchowski-Poisson (GSP) system
[see Eqs. (\ref{gsp1})-(\ref{gsp2}) of this paper] including an
arbitrary barotropic equation of state $P(\rho)$. This model has been
introduced by Chavanis in \cite{gen}. The GSP system can be viewed as
a generalized mean field Fokker-Planck equation (for a review of
nonlinear Fokker-Planck equations, see
\cite{frank,gfp}). It can be obtained from generalized stochastic
processes and it is associated with a notion of effective generalized
thermodynamics (E.G.T).  These equations can also provide a
generalized Keller-Segel (GKS) model of chemotaxis with a density
dependent diffusion coefficient
\cite{gfp}. For an isothermal equation of state $P=\rho k_{B}T/m$, we
recover the standard SP system and KS model (with appropriate
notations). Apart from the isothermal equation of state, the GSP
system and GKS model have been studied for: (i) a polytropic equation
of state $P=K\rho^{\gamma}$ \cite{lang} (ii) a logotropic equation of
state $P=A \ln\rho$ \cite{logotropes} (iii) a Fermi-Dirac equation of
state $P=P_{F.D.}(\rho)$ \cite{crrs,bln} (iv) an equation of state
$P=-T\rho_{max}
\ln(1-\rho/\rho_{max})$ taking into account
excluded volume effects \cite{degrad}. These are standard equations
of state introduced in astrophysics and statistical mechanics so that
it is natural to consider these equations of state in connexion to the
GSP system and GKS model.

Specializing on the polytropic equation of state $P=K\rho^{\gamma}$
with $\gamma=1+1/n$ \cite{lang}, the steady states of the GSP system correspond
to polytropic distributions associated with the Tsallis statistics
\cite{tsallis}. When coupled to the Poisson equation, we obtain
density profiles similar to {\it polytropic stars} in astrophysics
\cite{emden,chandrab}. For $d\ge 2$, there exists a critical index
$\gamma_{4/3}=2(d-1)/d$, i.e. $n_{3}=d/(d-2)$ \cite{lang}. For
$0<n<n_3$, the GSP system relaxes towards a stable steady state with a
compact support, similar to a classical white dwarf star (classical
white dwarf stars are equivalent to polytropes with index $n=3/2$ in
$d=3$ \cite{fowler}). For $n>n_3$, there is no stable equilibrium in
an unbounded domain so that the system can either collapse or
evaporate (see Fig. \ref{evap3} for an illustration). These different
regimes have been studied in
\cite{lang}. For $n=n_3$, the dynamics is critical. At this index,
there exists a critical mass $M_c(d)$ (for a given polytropic
constant $K$) \cite{lang} which is connected to the Chandrasekhar
mass of relativistic white dwarf stars (ultra-relativistic white
dwarf stars are equivalent to polytropes with index $n=3$ in $d=3$
\cite{chandra}). The object of the present paper is to study
numerically and, when possible, analytically this critical dynamics.
For $M<M_c$, we find that the system evaporates and we construct a
self-similar solution. For $M>M_c$, we find that the system
collapses. In a finite time $t_{coll}$, it forms a Dirac peak with
mass $M_c$ surrounded by a halo that has a pseudo self-similar
evolution. For $d=2$, the critical index $n_3\rightarrow +\infty$ so
that we recover the case of isothermal spheres whose dynamics is
known to be critical in $d=2$ \cite{sc}.

When we apply this model in the context of chemotaxis \cite{csmasse},
we find the existence of a critical mass $M_{c}(d)$ at the critical
index $n_{3}=d/(d-2)$. For $d=2$, we recover the well-known result
$M_{c}(d=2)=8\pi$ obtained within the standard Keller-Segel model (see
\cite{mt} and references therein) and for $d=3$, the critical mass
associated with the GKS model is $M_{c}(d=3)=202.8956...$ (in usual
dimensionless variables). This is similar to the Chandrasekhar
limiting mass of white dwarf stars. The existence of a limiting mass
for bacterial populations at the critical index $n_{3}$ and its
connexion to the Chandrasekhar mass was pointed out in
\cite{wd,csmasse} (and implicitly in \cite{lang}).
This is another illustration of the numerous analogies that exist
between self-gravitating systems and bacterial populations
\cite{crrs}.

The paper is organized as follows. In Sec. \ref{sec_wdp}, we briefly
recall the connexion between white dwarf stars and gaseous
polytropes. In Sec. \ref{sec_sglp}, we recall the basic properties of
the SP and GSP systems and describe the behavior of the solutions
depending on the index $n$ and the dimension of space $d$. As the
problem is very rich, involving many different cases ($\sim 30$), a summary of
previously obtained results, completed by new results and new
discussion, is required to understand the place of the present study
in the general problem (see also Tables \ref{Table1} and
\ref{Table2} for an overview). Then, we consider more specifically
the particular index $n=n_{3}$ which presents a critical dynamics that
was mentioned, but not studied, in our previous paper \cite{lang}.  In
Sec. \ref{sec_dim}, we show that this critical value can be understood
from a simple dimensional analysis.  In Sec.
\ref{sec_collapse}, we study the critical collapse dynamics and
extend the results obtained in $d=2$ for isothermal ($n=+\infty$)
systems \cite{sc} to the case of {\it critical polytropes}
($n=n_{3}$) in $d>2$. In Sec. \ref{sec_evaporation}, we study the
evaporation dynamics in unbounded space. We show that for $n>n_3$,
self-gravity becomes negligible for large times so that the
evaporation is eventually controlled by the pure (anomalous)
diffusion. For $n=n_3$, gravity remains relevant at any time so that
there exists a self-similar solution for which all the terms of the
GSP system scale the same way. Finally, in Sec. \ref{sec_analogy},
we transpose our main results to the context of chemotaxis using
notations and parameters adapted to this problem (this is to
facilitate the comparison with the results obtained in mathematical
biology).

Our numerical  and analytical study  was  conducted  in parallel to  a
mathematical work by Blanchet {\it et al.}
\cite{bcl} who obtained rigorous results for the critical dynamics of
the GSP system and GKS model introduced in our paper
\cite{lang}. These two independent studies have different motivations
and use very different methods so they are complementary to each
other.

\section{White dwarf stars and polytropes}
\label{sec_wdp}

In this section, we briefly recall the connexion between the maximum
mass of white dwarf stars (Chandrasekhar's mass
\cite{chandra}) and the theory of self-gravitating polytropic
spheres \cite{emden,chandrab}.

In simplest models of stellar structure, a white dwarf star can be
viewed as a degenerate gas sphere in hydrostatic equilibrium. The
pressure is entirely due to the quantum pressure of the electrons
(resulting from Pauli's exclusion principle for fermions) while the
density of the star is dominated by the mass of the protons. The condition of
hydrostatic equilibrium coupled to the Poisson equation reads
\begin{equation}
\nabla P=-\rho\nabla \Phi,\qquad \Delta\Phi=4\pi G\rho,
\label{wdp1}
\end{equation}
and the equation of state of a degenerate gas of relativistic fermions
at $T=0$ can be written parametrically as follows \cite{chandraM}
\begin{equation}
P=A_{2}f(x), \qquad \rho=Bx^{3},
\label{wdp2}
\end{equation}
where
\begin{equation}
A_{2}={\pi m^{4}c^{5}\over 3h^{3}}, \qquad
B={8\pi m^{3}c^{3}\mu H\over 3h^{3}}, \label{wdp3}
\end{equation}
\begin{eqnarray}
f(x)=x(2x^{2}-3)(1+x^{2})^{1/2}+3\ {\sinh}^{-1}x,
\label{wdp4}
\end{eqnarray}
where $m$ is the mass of the electrons, $H$ is the mass of the protons
and $\mu$ is the molecular weight.  The function $f(x)$ has the
asymptotic behaviors $f(x)\simeq (8/5) x^{5}$ for $x\ll 1$ and
$f(x)\simeq 2 x^{4}$ for $x\gg 1$.  The classical limit corresponds to
$x\ll 1$ and the ultra-relativistic limit to $x\gg 1$. In these
limits, the white dwarf star is equivalent to a polytropic gas sphere
with an equation of state $P=K\rho^{\gamma}$. The index $n$ of the
polytrope is defined by $\gamma=1+1/n$. In $d=3$ dimensions,
polytropes are self-confined for $n<5$ and they are stable (with
respect to the Euler-Poisson system) for $n\le 3$ (for $n=3$ they are
marginally stable). The mass-radius relation is given by \cite{chandrab}:
\begin{eqnarray}
M^{(n-1)/n}R^{(3-n)/n}=\frac{K(1+n)}{G(4\pi)^{1/n}}\omega_{n}^{(n-1)/n},
\label{wdp5}
\end{eqnarray}
where $\omega_{n}$ is a constant (depending only on the index $n$ of
the polytrope) that can be expressed in terms of the solution of the
Lane-Emden equation \cite{emden}.

\begin{figure}[htbp]
\centerline{
\includegraphics[width=8cm,angle=0]{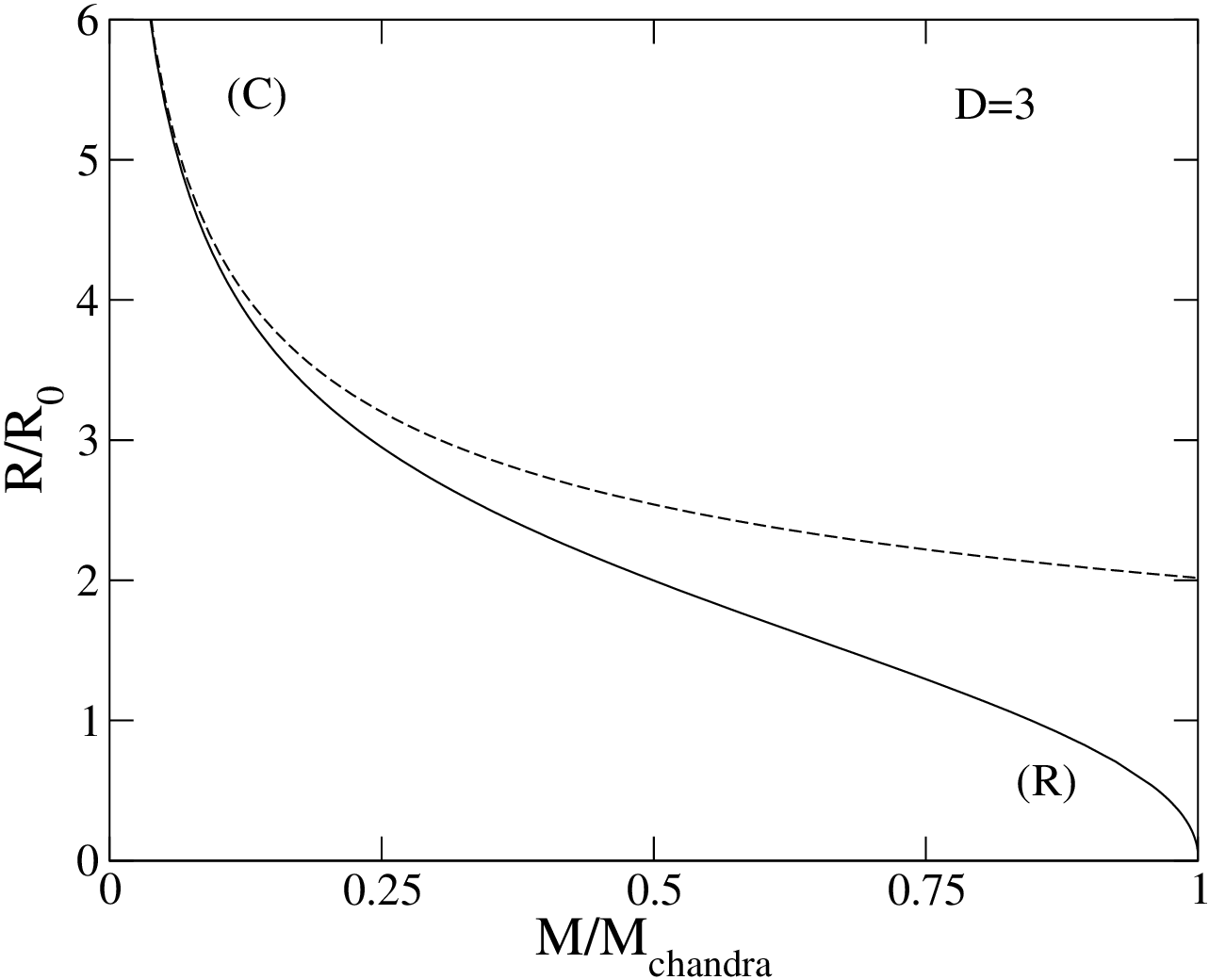}
} \caption[]{Mass-radius relation for relativistic white dwarf stars
at $T=0$ \cite{chandra}.  The radius vanishes for a limiting mass
$M_{Chandra}$ corresponding to the ultra-relativistic limit (R). The
dashed line corresponds to the classical limit (C).} \label{chandra3}
\end{figure}

\begin{figure}[htbp]
\centerline{
\includegraphics[width=8cm,angle=0]{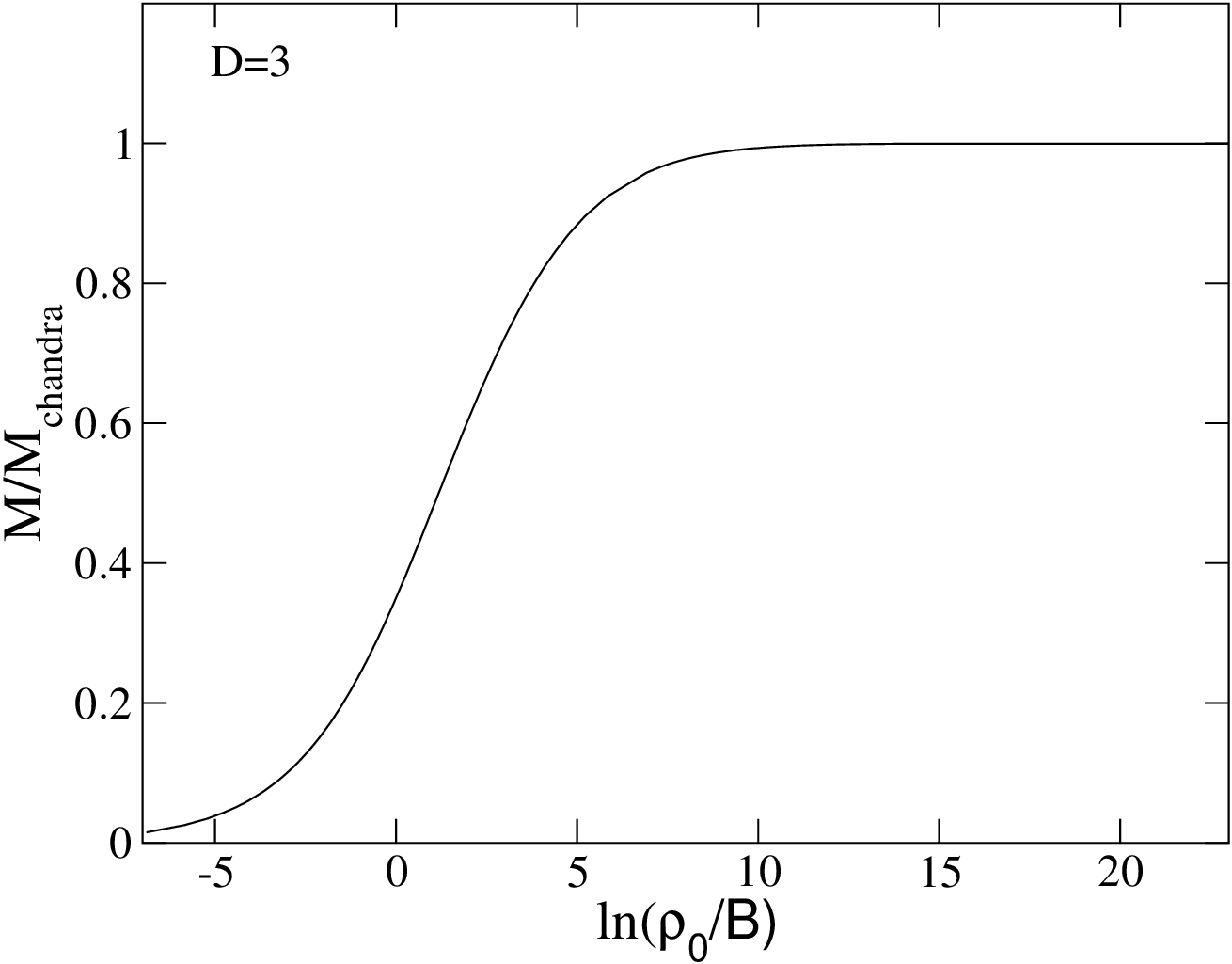}
} \caption[]{Mass versus central density for relativistic white dwarf
stars at $T=0$. Equilibrium states only exist for $M<M_{Chandra}$. For
$M=M_{Chandra}$, the density profile is a Dirac peak. For
$M>M_{Chandra}$, the system is expected to collapse and form a neutron
star or a black hole. The corresponding density
profiles are represented in Fig. 4 of \cite{chandraM}.}
\label{massedensiteD3}
\end{figure}

In the classical case $x\ll 1$, the equation of state
takes the form
\begin{equation}
P=K_{1}\rho^{5/3},
\label{wdp6}
\end{equation}
with
\begin{equation}
 K_{1}={1\over 5}\biggl ({3\over 8\pi}\biggr )^{2/3} {h^{2}\over m
 (\mu H)^{5/3}}.
\label{wdp7}
\end{equation}
Therefore a classical white dwarf star is equivalent to a polytrope of
index  $n=3/2$. The mass-radius relation is given by
\begin{equation}
M^{1/3}R={1\over 2}\biggl ({3\over 32 \pi^{2}}\biggr
)^{2/3}{h^{2}\over mG(\mu H)^{5/3}}\ \omega_{3/2}^{1/3},
\label{wdp8}
\end{equation}
with $\omega_{3/2}=132.3843...$. It exhibits the familiar
$MR^{3}\sim 1$ scaling.

In the ultra-relativistic limit $x\gg 1$, the equation of state takes
the form
\begin{equation}
P=K_{2}\rho^{4/3},
\label{wdp9}
\end{equation}
with
\begin{equation}
 K_{2}={1\over 4}\biggl ({3\over 8\pi}\biggr )^{1/3}{hc\over (\mu H)^{4/3}}.
\label{wdp10}
\end{equation}
Therefore, an ultra-relativistic white dwarf star is equivalent to a
polytrope of index $n=3$. For this index, the relation (\ref{wdp5})
leads to a unique value of the mass
\begin{equation}
M_c=\biggl ({3\over 32\pi^{2}}\biggr )^{1/2}\omega_{3}\biggl ({hc\over
G}\biggr )^{3/2}{1\over (\mu H)^{2}},
\label{wdp11}
\end{equation}
with $\omega_{3}=2.01824...$. This is the Chandrasekhar mass
\begin{equation}
M_c=0.196701...\biggl ({hc\over G}\biggr )^{3/2}{1\over (\mu
H)^{2}}\simeq 5.76 M_{\odot}/\mu^{2}.
\label{wdp12}
\end{equation}
Considering the general mass-radius relation of partially relativistic
white dwarf stars (see Fig. \ref{chandra3}), we note that, for this
limiting value, the radius $R$ of the configuration tends to
zero. This leads to a Dirac peak with mass $M_c$. Thus, the
Chandrasekhar mass represents the maximum mass of white dwarf stars
(see Fig. \ref{massedensiteD3}). There is no hydrostatic equilibrium
configuration for $M>M_{c}$.

If we extend Chandrasekhar's theory to a $d$-dimensional universe
\cite{wd}, we find that white dwarf stars become unstable in a
universe with $d\ge 4$ dimensions (in $d=4$, classical white dwarf
stars exist for a unique value of the mass
$M=M_c=0.0143958...h^4/(m^2G^2\mu^3 H^3)$ and they are
marginally stable). Therefore, the dimension $d=3$ of our universe is
very special regarding the laws of gravity. This is the largest
dimension of space at which all the series of equilibria of white
dwarf stars (from classical to ultra-relativistic) is stable. This may
have implications regarding the Anthropic
Principle.

\section{Self-gravitating Langevin particles}
\label{sec_sglp}

\subsection{The generalized Smoluchowski-Poisson system}
\label{sec_gsp}

In this paper, we shall study a dynamical model of self-gravitating
systems whose steady states reproduce the condition of hydrostatic
equilibrium Eq. (\ref{wdp1}). Specifically, we consider the generalized
Smoluchowski-Poisson system \cite{gen}:
\begin{equation}
\frac{\partial\rho}{\partial t}=\nabla\cdot \left\lbrack
\frac{1}{\xi}\left (\nabla P+\rho\nabla\Phi\right )\right\rbrack,
\label{gsp1}
\end{equation}
\begin{equation}
\Delta\Phi=S_{d}G\rho,
\label{gsp2}
\end{equation}
where $P(\rho)$ is a barotropic equation of state, i.e.  the pressure
$P({\bf r},t)$ depends only on the density of particles $\rho({\bf
r},t)$. This model describes a {\it dissipative} gas of
self-gravitating Langevin particles in an overdamped limit
$\xi\rightarrow +\infty$ (where inertial effects are neglected) and in
the thermodynamic limit $N\rightarrow +\infty$ (where the mean field
approximation becomes exact) \cite{virial2,lemou}.  The GSP system is
a particular example of generalized mean field Fokker-Planck equation
\cite{gfp}. It is associated to   a stochastic process of the form
\begin{equation}
\frac{d{\bf r}}{dt}=-\frac{1}{\xi}\nabla \Phi+\sqrt{\frac{2
P(\rho)}{\rho\xi}}{\bf R}(t),\label{gsp3}
\end{equation}
where ${\bf R}(t)$ is a white noise with $\langle {\bf
R}(t)\rangle={\bf 0}$ and $\langle
R_i(t)R_j(t')\rangle=\delta_{ij}\delta(t-t')$. This stochastic process
describes the evolution of each of the $N$ Langevin particles
interacting through the mean field potential $\Phi({\bf r},t)$. For
sake of generality, we have allowed the strength of the noise term in
Eq. (\ref{gsp3}) to depend on the local distribution of particles.
This gives rise to anomalous diffusion and generalized pressure laws
as discussed in
\cite{frank,gfp}.

The Lyapunov functional (or generalized free energy) associated with the GSP system is
\begin{equation}
F=\int\rho\int^{\rho}\frac{P(\rho')}{\rho'^{2}}\,d\rho' d{\bf
r}+\frac{1}{2}\int\rho\Phi\, d{\bf r}.
\label{gsp4}
\end{equation}
Easy calculations lead to
\begin{equation}
\dot F=-\int\frac{\xi}{\rho}(\nabla P+\rho\nabla\Phi)^{2}d{\bf r}\le 0.
\label{gsp5}
\end{equation}
The GSP system has the following properties: (i) the total mass is
conserved. (ii) $\dot F\le 0$. (iii) $\dot F=0$ $\Leftrightarrow$
$\nabla P+\rho\nabla\Phi={\bf 0}$ (hydrostatic equilibrium)
$\Leftrightarrow$ $\partial_{t}\rho=0$. (iv) $\rho_{eq}({\bf r})$ is a
steady state of the GSP system iff it is a critical point of $F[\rho]$
at fixed mass. (v) A steady state of the GSP system is linearly
dynamically stable iff it is a (local) minimum of $F[\rho]$ at fixed
mass \footnote{Since the free energy $F[\rho]$ coincides with the
energy functional ${\cal W}[\rho]$ of a barotropic gas (up to a
positive macroscopic kinetic term) \cite{aaantonov}, we conclude that
$\rho_{eq}({\bf r})$ is linearly dynamically stable with respect to
the GSP system iff it is formally nonlinearly dynamically stable with
respect to the barotropic Euler-Poisson system
\cite{aaantonov}.}. By Lyapunov's direct
method \cite{frank}, we know that if $F[\rho]$ is bounded from below,
the GSP system will relax towards a (local) minimum of $F[\rho]$ at
fixed mass for $t\rightarrow +\infty$. If $F[\rho]$ has several
minima, the choice of the selected minimum will depend on a notion of
basin of attraction: if the initial condition is sufficiently ``close'' to
the minimum $\rho_{eq}({\bf r})$, the distribution $\rho({\bf r},t)$
will converge towards $\rho_{eq}({\bf r})$ for $t\rightarrow
+\infty$. Finally, if $F[\rho]$ has no global minimum (as can be the
case for self-gravitating systems), the system can either tend to a
local minimum (metastable) if it exists, or undergo collapse or
evaporation.

We are not claiming that this simple model accurately describes the
dynamics of white dwarf stars or other astrophysical systems. However,
we have undertaken a systematic study of the GSP system for different
equations of state that have been considered in astrophysics. The main
interest of this model is its simplicity (while being still very rich)
which enables an accurate numerical and analytical treatment. This can
be viewed as a first step before considering other, more realistic,
dynamical models of self-gravitating systems. On the other hand, in a
completely different context, this model is isomorphic to the standard
Keller-Segel model describing the chemotaxis of bacterial populations
(see Sec. \ref{sec_analogy}). This is a further motivation to study
this type of equations at a general level \footnote{We shall study the
problem in $d$ dimensions because: (i) We have found that the structure
of the mathematical problem with the dimension of space is very rich
\cite{sc,lang,wd}, exhibiting several characteristic dimensions. (ii)
In gravity, the usual dimension is $d=3$ but, in biology (see
Sec. \ref{sec_analogy}), the bacteria (or cells) are compelled to lie
on a plane so that $d=2$.}.

\subsection{Isothermal spheres}
\label{sec_i}

For an isothermal equation of state $P=\rho
k_{B}T/m$, we recover the standard Smoluchowski-Poisson system \cite{crs}:
\begin{equation}
\frac{\partial\rho}{\partial t}=\nabla\cdot \left\lbrack
\frac{1}{\xi}\left (\frac{k_{B}T}{m}\nabla \rho+\rho\nabla\Phi\right
)\right\rbrack,
\label{i1}
\end{equation}
\begin{equation}
\Delta\Phi=S_{d}G\rho.
\label{i2}
\end{equation}
Equation (\ref{i1}) is an ordinary
mean-field Fokker-Planck equation associated with a Langevin dynamics
of the form
\begin{eqnarray}
\label{i3}
\frac{d{\bf r}}{dt}=-\frac{1}{\xi}\nabla \Phi+\sqrt{\frac{2k_{B}T}{\xi m}}{\bf R}(t),
\end{eqnarray}
where the strength of the noise is constant.
The Lyapunov functional of the SP system can be written
\begin{equation}
F=k_{B}T\int \frac{\rho}{m}\ln\frac{\rho}{m} \,d{\bf
r}+\frac{1}{2}\int\rho\Phi \,d{\bf r}.
\label{i4}
\end{equation}
This is the Boltzmann free energy $F_B=E-TS_B$ where $E=(1/2)\int
\rho\Phi \,d{\bf r}$ is the energy and $S_B=-k_{B}\int ({\rho}/{m})\ln
({\rho}/{m}) \,d{\bf r}$ is the Boltzmann entropy.  The stationary
solutions of the SP system are given by the Boltzmann distribution
\begin{equation}
\rho=A e^{-\beta m \Phi},
\label{i5}
\end{equation}
where $A$ is a constant determined by the mass $M$.  These steady
states can also be obtained by extremizing $F$ at fixed mass, writing
$\delta F-\alpha\delta M=0$, where $\alpha$ is a Lagrange
multiplier. The equilibrium distribution is obtained by substituting
Eq. (\ref{i5}) into Eq. (\ref{i2}) leading to the Boltzmann-Poisson
equation. Specializing on spherically symmetric distributions and
defining
\begin{equation}
\rho=\rho_{0}e^{-\psi(\xi)}, \quad \xi=r/r_{0}=(S_{d}\beta Gm\rho_0)^{1/2}r,
\label{i6}
\end{equation}
where $\rho_0$ is the central density, we find after simple algebra
that $\psi$ is solution of the Emden equation
\begin{equation}
\frac{1}{\xi^{d-1}}\frac{d}{d\xi}\left
(\xi^{d-1}\frac{d\psi}{d\xi}\right )=e^{-\psi},
\label{i7}
\end{equation}
with $\psi=0$ and $\psi'=0$ at $\xi=0$. The Emden equation can also be
obtained from the fundamental equation of hydrostatic equilibrium for
an isothermal equation of state \cite{emden,chandrab,sc}. Note that the
isothermal spheres have a self-similar structure
$\rho(r)/\rho_{0}=e^{-\psi(r/r_0)}$: if we rescale the central density
and the radius appropriately, they have the same profile
$e^{-\psi(\xi)}$. This property is called homology \cite{chandrab}.

For $d=1$, the SP system is equivalent to the Burgers equation
\cite{acedo,mt} and it relaxes towards the Camm distribution \cite{camm}
which is a global minimum of free energy for any temperature.  For
$d>2$, there is no steady state with finite mass in an unbounded
domain because the density of an isothermal self-gravitating system
decreases as $\rho\sim r^{-2}$ for $r\rightarrow +\infty$
\cite{chandrab}.  We shall thus enclose the system within a box of radius
$R$ \footnote{It may appear artificial to put the system in a
``box''. In gravity, the box delimitates the region of space where the
system can be assumed isolated from the surrounding and where
statistical mechanics applies. In biology (see Sec. \ref{sec_analogy}),
the box has a physical meaning since it represents the container in
which the bacteria (or cells) are confined.}. For box-confined
systems, we must integrate the Emden equation (\ref{i7}) until the
normalized box radius $\xi=\alpha$ with
\begin{eqnarray}
\alpha=(S_{d}\beta Gm\rho_0)^{1/2}R.
\label{i8}
\end{eqnarray}
It is useful to define a
dimensionless control parameter
\begin{eqnarray}
\eta=\frac{\beta GMm}{R^{d-2}}.
\label{i9}
\end{eqnarray}
Using the conservation of mass or the Gauss theorem, we get \cite{sc}:
\begin{eqnarray}
\eta=\alpha\psi'(\alpha).
\label{i10}
\end{eqnarray}
This equation relates the central density to the mass and the
temperature. More precisely, the relation $\eta(\alpha)$ gives the
mass $M$ as a function of the central density (for a fixed
temperature $T$) or the temperature $T$ as a function of the density
contrast ${\cal R}\equiv \rho(0)/\rho(R)=e^{\psi(\alpha)}$ (for a
fixed mass $M$). The curve $\eta(\alpha)$ is plotted in Fig. 3 of
\cite{sc}. For $2<d<10$, the series of equilibria $\eta(\alpha)$
oscillates and presents a first turning point at
$\eta_{c}=\eta(\alpha_{1})$ (for $d\ge 10$, the series of equilibria
does not display any oscillation).  According to Poincar\'e's turning
point argument \cite{katz,ijmpb}, configurations with
$\alpha>\alpha_{1}$ are unstable (saddle points of free energy at
fixed mass). This concerns in particular the singular isothermal
sphere corresponding to $\alpha\rightarrow +\infty$.  Configurations
with $\alpha<\alpha_{1}$ are metastable (local minima of free energy
at fixed mass) and they exist only for $\eta\le\eta_c$. There is no
global minimum of free energy for self-gravitating isothermal
spheres. For $\eta\le
\eta_{c}$, depending on the form of the initial density profile, the
SP system can either relax towards a box-confined isothermal sphere
(metastable) or collapse. This behavior has been illustrated
numerically in Fig. 16 of \cite{crs}.  For $\eta>\eta_{c}$ the SP
system undergoes gravitational collapse.  This self-similar collapse,
followed by the formation of a Dirac peak, has been studied in detail
in \cite{sc,post}.  If we remove the box, the SP system can either
collapse or evaporate depending on the initial condition (this
behavior will be illustrated numerically in Sec. \ref{sec_sup}).

The dimension $d=2$ is critical and has been studied in detail in
\cite{sc,virial1}. The solution of the Emden equation is known analytically \cite{ostriker}:
\begin{equation}
e^{-\psi}=\frac{1}{\left (1+\frac{1}{8}\xi^{2}\right )^{2}}.
\label{i11}
\end{equation}
In an unbounded domain, the density profile extends to infinity but
the total mass is finite because the density decreases as $r^{-4}$ for
$r\rightarrow +\infty$. The total mass $M=\int_{0}^{+\infty}\rho 2\pi r dr$
is given by
\begin{equation}
M=\frac{1}{\beta Gm}\int_{0}^{+\infty} e^{-\psi}\xi
d\xi=\frac{1}{\beta Gm}\lim_{\xi\rightarrow +\infty} \xi \psi'(\xi),
\label{i12}
\end{equation}
where we have used the Emden equation (\ref{i7}) to get the last
equality. Using Eq. (\ref{i11}), we find that $\xi\psi'\rightarrow 4$
for $\xi\rightarrow +\infty$. This yields a unique value of the mass
(for a fixed temperature), or equivalently a unique value of the
temperature (for a fixed mass) given by
\begin{eqnarray}
M_{c}=\frac{4k_{B}T}{Gm}, \qquad k_{B}T_{c}=\frac{GMm}{4}.
\label{i13}
\end{eqnarray}
For $T=T_{c}$ or $M=M_{c}$, we have an infinite  family of steady states
\begin{eqnarray}
\rho(r)=\frac{\rho_{0}}{\left (1+\frac{1}{8}(r/r_0)^2\right
)^2},\quad \rho_{0}r_0^2=\frac{k_{B}T}{2\pi Gm}, \label{i14}
\end{eqnarray}
parameterized by the central density $\rho_{0}$. For
$\rho_{0}\rightarrow +\infty$, we obtain a Dirac peak with mass
$M_c$. The steady states (\ref{i14}) have the same value of the free
energy, independently on the central density $\rho_{0}$ (see Appendix
\ref{sec_virial}) and they are marginally stable ($\delta^{2}F=0$). For  $T\neq T_{c}$ or $M\neq M_c$, there is no steady state in an infinite domain.
For $T>T_c$ or $M<M_c$, the solution of the SP system evaporates and
for $T<T_c$ or $M>M_c$, the solution of the SP
system collapses. These different regimes have been discussed in
detail in \cite{sc,virial1}.

If we consider box confined configurations in $d=2$, we observe that
the control parameter (\ref{i9}) is independent on the box radius
and can be written
\begin{eqnarray}
\eta=\beta GMm=4\frac{M}{M_{c}}=4 \frac{T_{c}}{T}.
\label{i15}
\end{eqnarray}
Using Eqs. (\ref{i10}) and (\ref{i11}), we obtain the relation
$\eta(\alpha)=(\alpha^2/2)/(1+\alpha^2/8)$ between the central
density, the mass and the temperature. The density profiles are given
by Eq. (\ref{i14}) with $8(r_0/R)^2=(T/T_c-1)=(M_c/M-1)$ so the
central density is now determined by the mass $M$ or the temperature
$T$. Equilibrium states exist only for $\eta\le \eta_{c}=4$, 
i.e. $M\le M_{c}$ or $T\ge T_{c}$ and, since the series of 
equilibria is monotonic, they are fully
stable (global minima of free energy at fixed mass). In that case, the
SP system tends to a box-confined isothermal sphere. For
$\eta=\eta_{c}=4$, i.e. $M=M_{c}$ or $T=T_{c}$, the steady state is a
Dirac peak containing all the mass.  For $\eta>\eta_{c}=4$ the SP
system undergoes gravitational collapse (see Sec. \ref{sec_collapse}).

\begin{figure}[htbp]
\centerline{
\includegraphics[width=8cm,angle=0]{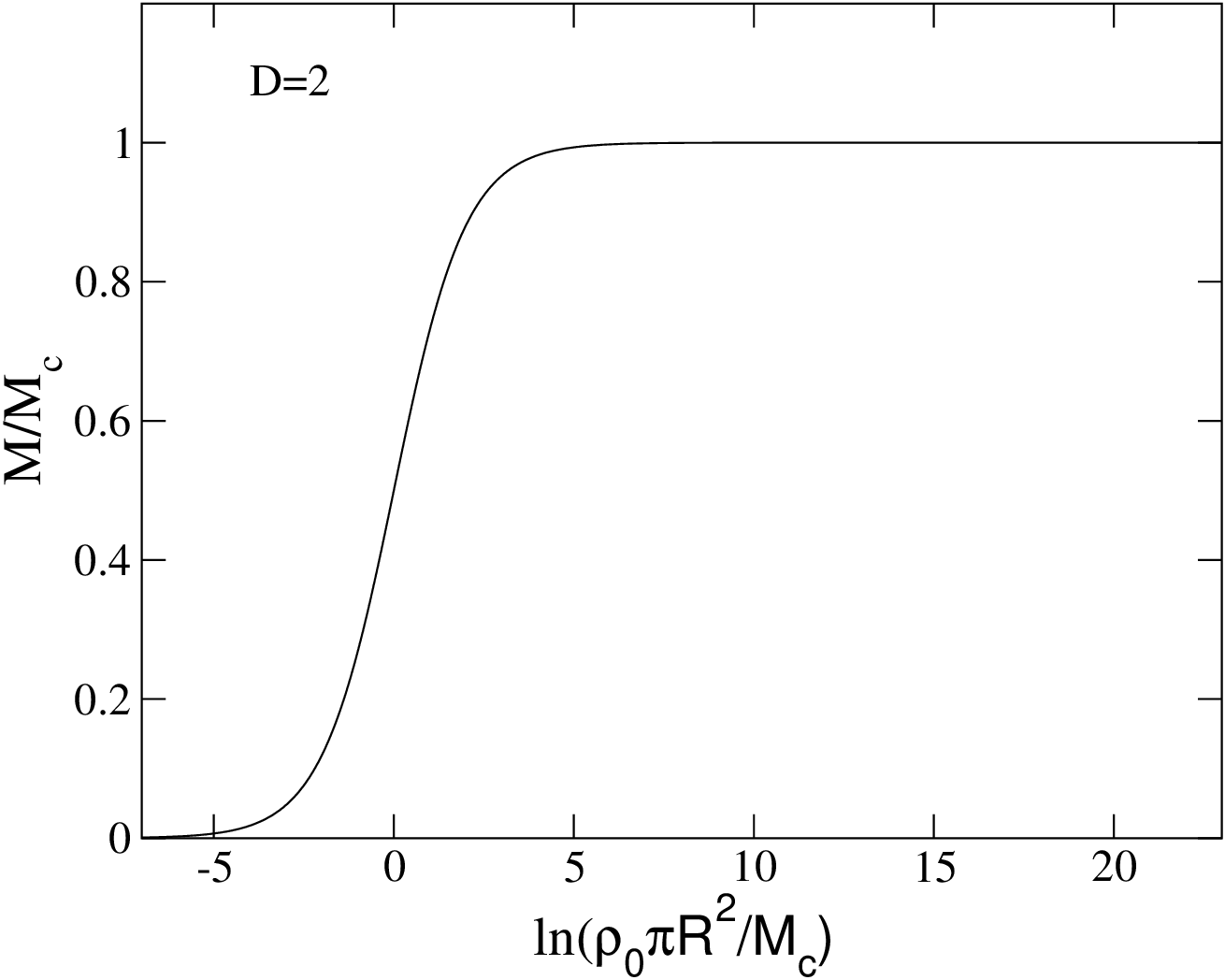}
} \caption[]{Mass as a function of the central density for
two-dimensional box-confined self-gravitating isothermal spheres with
fixed temperature. Equilibrium states exist only for $M\le M_{c}$. For
$M=M_{c}$, the density profile is a Dirac peak and for $M>M_c$ the
system undergoes gravitational collapse. More precisely, this curve
represents $\eta(\alpha)/4$ so it also gives the inverse temperature
$T_{c}/T$ as a function of the density contrast ${\cal
R}=\rho_{0}/\rho(R)={\cal R}(\alpha)$ for a fixed
mass. The corresponding density profiles are
represented in Fig. 1 of \cite{mt}. }
\label{mrho}
\end{figure}

The mass-central density (for a fixed temperature) of two-dimensional
isothermal spheres is plotted in Fig. \ref{mrho}. We note the striking
analogy with the mass-central density of white dwarf stars in
Fig. \ref{massedensiteD3}. Therefore, the critical mass (\ref{i13}) of
isothermal spheres in two dimensions shares some resemblance with the
Chandrasekhar mass. We shall show in the next section that this
analogy (which is not obvious a priori) bears more significance than
is apparent at first sight.

\subsection{Complete polytropes}
\label{sec_up}

If we consider a polytropic equation of state $P=K\rho^{\gamma}$ with
$\gamma=1+1/n$, we get the polytropic Smoluchowski-Poisson system \cite{lang}:
\begin{equation}
\frac{\partial\rho}{\partial t}=\nabla\cdot \left\lbrack
\frac{1}{\xi}\left (K\nabla \rho^{\gamma}+\rho\nabla\Phi\right
)\right\rbrack,
\label{up1}
\end{equation}
\begin{equation}
\Delta\Phi=S_{d}G\rho.
\label{up2}
\end{equation}
Equation (\ref{up1}) is a generalized mean field Fokker-Planck equation
associated with the stochastic process
\begin{eqnarray}
\label{up3} \frac{d{\bf r}}{dt}=-\frac{1}{\xi}\nabla
\Phi+\sqrt{\frac{2K}{\xi}}\rho^{(\gamma-1)/2}{\bf R}(t),
\end{eqnarray}
where the strength of the noise depends on the local density as a power law
\cite{borland}.  The Lyapunov functional of the polytropic SP system can be written
\begin{equation}
F=\frac{K}{\gamma -1}\int (\rho^{\gamma}-\rho) \,d{\bf
r}+\frac{1}{2}\int\rho\Phi \,d{\bf r}.
\label{up4}
\end{equation}
It can be interpreted as a generalized free energy of the form
$F=E-T_{eff} S$ where $E=(1/2)\int \rho\Phi \,d{\bf r}$ is the energy,
$T_{eff}=K$ is an effective temperature (polytropic temperature) and
$S=-1/(\gamma-1)\int (\rho^{\gamma}-\rho)\,d{\bf r}$ is the Tsallis
entropy (the polytropic index $\gamma$ plays the role of the Tsallis
$q$ parameter). For $\gamma=1$, i.e. $n\rightarrow +\infty$, the
polytropic equation of state $P=K\rho^{\gamma}$ reduces to
$P=K\rho$. It coincides with an isothermal equation of state $P=\rho
k_{B}T/m$ with temperature $K=k_{B}T/m$ leading to the standard
Smoluchowski-Poisson system (\ref{i1})-(\ref{i2}).

The stationary solutions of the GSP system (\ref{up1}) are
given by the Tsallis distributions
\begin{equation}
\rho=\left \lbrack
\lambda-\frac{\gamma-1}{K\gamma}\Phi\right\rbrack_{+}^{1/(\gamma-1)},
\label{up5}
\end{equation}
where $\lambda$ is a constant determined by the mass $M$ (by
definition $[x]_{+}=x$ if $x\ge 0$ and $[x]_{+}=0$ if $x<0$). These
steady states can also be obtained by extremizing $F$ at fixed mass,
writing $\delta F-\alpha\delta M=0$, where $\alpha$ is a Lagrange
multiplier. The equilibrium distribution is obtained by substituting
Eq. (\ref{up5}) into Eq. (\ref{up2}) leading to the Tsallis-Poisson
equation. Specializing on spherically symmetric solutions and defining
\begin{equation}
\rho=\rho_0\theta^{n}(\xi), \quad \xi=r/r_{0}, \quad r_0=\left\lbrack
\frac{K(1+n)}{S_{d}G\rho_0^{1-1/n}}\right\rbrack^{1/2},
\label{up6}
\end{equation}
where $\rho_0$ is the central density, we find after
simple algebra that $\theta$ is solution of the Lane-Emden equation
\begin{equation}
\frac{1}{\xi^{d-1}}\frac{d}{d\xi}\left
(\xi^{d-1}\frac{d\theta}{d\xi}\right )=-\theta^{n},
\label{up7}
\end{equation}
with $\theta=1$ and $\theta'=0$ at $\xi=0$. The Lane-Emden equation
can equivalently be derived from the fundamental equation of
hydrostatic equilibrium with a polytropic equation of state
\cite{emden,chandrab,lang}. Note that the polytropic spheres have a self-similar
structure $\rho(r)/\rho_{0}=\theta^{n}(r/r_0)$: if we rescale the
central density and the radius appropriately, they have the same
profile $\theta^{n}(\xi)$. This property is called homology \cite{chandrab}.

In this paper, we restrict ourselves to $n>0$. Let us first discuss
the case $d>2$. For $n>n_{5}=(d+2)/(d-2)$, unbounded self-gravitating
polytropes have infinite mass because their density profile decreases
like $r^{-\alpha}$ for $r\rightarrow +\infty$, with
$\alpha=2n/(n-1)$. For $n<n_{5}=(d+2)/(d-2)$, they are
self-confined. In that case, the function $\theta$ vanishes at
$\xi=\xi_{1}$ and the density vanishes at $R_{*}=r_{0}\xi_{1}$ which
defines the radius of the polytrope.  The relation between the radius
and the central density is
\begin{equation}
R_{*}=\left\lbrack
\frac{K(1+n)}{S_{d}G\rho_0^{1-1/n}}\right\rbrack^{1/2}\xi_{1}.
\label{up8}
\end{equation}
The total mass $M=\int_{0}^{R_{*}}\rho S_{d}r^{d-1}dr$ can be written as
\begin{equation}
M=S_{d}\rho_0
r_{0}^{d}\int_{0}^{\xi_{1}}\theta^{n}\xi^{d-1}d\xi=-S_{d}\rho_0
r_{0}^{d}\xi_{1}^{d-1}\theta'_{1}, \label{up9}
\end{equation}
where we have used the Lane-Emden equation (\ref{up7}) to get the last
equality. Therefore, the relation between the mass and the central
density is
\begin{equation}
M=-S_{d}\rho_0\left\lbrack \frac{K(1+n)}{S_{d}G\rho_0^{1-1/n}}
\right\rbrack^{d/2}\xi_{1}^{d-1}\theta'_{1}.
\label{up10}
\end{equation}
Eliminating the central density between Eqs. (\ref{up8}) and
(\ref{up10}) and introducing the index
\begin{equation}
n_{3}=\frac{d}{d-2},
\label{up11}
\end{equation}
we get the mass-radius relation
\begin{eqnarray}
M^{(n-1)/n}R_{*}^{\lbrack (d-2)(n_3-n)
\rbrack/n}=\frac{K(1+n)}{GS_{d}^{1/n}}\omega_{n}^{(n-1)/n},
\label{up12}
\end{eqnarray}
where
\begin{eqnarray}
\omega_{n}=-\xi_{1}^{(n+1)/(n-1)}\theta'_{1}.
\label{up13}
\end{eqnarray}

Let us introduce the polytropic temperature
\begin{eqnarray}
\Theta=\frac{K(1+n)}{nS_{d}^{1/n}}.
\label{up14}
\end{eqnarray}
For $0<n<n_{3}$ there is one, and only one, steady state for each
mass $M$ and temperature $\Theta$ and it is fully stable (global
minimum of $F$ at fixed mass). The GSP system will relax towards
this complete polytrope (note that for $n=1$ the radius $R_{*}$ of
the polytrope is independent on the mass).  For $n_{3}<n<n_{5}$
there is one, and only one, steady state for each mass $M$ and
temperature $\Theta$ but it is unstable (saddle point of $F$ at
fixed mass). In that case, the system will either collapse or
evaporate. The index $n_{3}$ is {\it critical}. For $n=n_{3}$, there
exists steady solutions for a unique value of the mass (at fixed
temperature $\Theta$):
\begin{eqnarray}
M_{c}=\left (
\frac{n_{3}\Theta}{G}\right )^{n_3/(n_3-1)}\omega_{n_3},
\label{up15}
\end{eqnarray}
or for a unique temperature (at fixed mass $M$):
\begin{eqnarray}
\Theta_{c}=\frac{G}{n_3}\left (\frac{M}{\omega_{n_3}}\right )^{(n_3-1)/n_3}.
\label{up16}
\end{eqnarray}
For $d=3$, we have
$M_{c}=({3\Theta}/{G})^{3/2}\omega_{3}=10.487... ({\Theta}/{G})^{3/2}$
and
$\Theta_{c}=({G}/{3})({M}/{\omega_{3}})^{2/3}=0.20872... ({G}/{M})^{2/3}$. As
we have seen in Sec. \ref{sec_wdp}, the Chandrasekhar limiting mass of
relativistic white dwarf stars is connected to the limiting mass
(\ref{up15}) of critical polytropes.  For a polytropic equation of state with
critical index $n=n_3$, and for $M=M_c$, we get an infinite family of
steady solutions
\begin{eqnarray}
\rho(r)=\rho_{0}\theta^{n_3}(r/r_0), \quad
\rho_{0}r_{0}^{d}=\frac{1}{S_d}\left (\frac{\Theta n_{3}}{G}\right
)^{d/2}, \label{up17}
\end{eqnarray}
parameterized by the central density $\rho_{0}$. For
$\rho_{0}\rightarrow +\infty$, the density profile tends to a Dirac
peak with mass $M_{c}$. These solutions have the same equilibrium free energy
$F[\rho_{eq}]=-dKM/(d-2)$ independently on the central density
$\rho_{0}$ (see Appendix \ref{sec_virial}) and they  are marginally stable
($\delta^{2}F=0$).  For $M<M_{c}$ (at fixed
temperature) or $\Theta>\Theta_c$ (at fixed mass), the solutions of
the GSP system evaporate and for $M>M_{c}$ (at fixed temperature) or
$\Theta<\Theta_c$ (at fixed mass), they collapse. These different
regimes will be studied in detail in Secs. \ref{sec_collapse} and \ref{sec_evaporation}.

For $d=2$, we find that $n_{3}\rightarrow +\infty$, so we realize that
isothermal systems ($n=+\infty$) in two dimensions are similar to
critical polytropes ($n=n_{3}$) in higher dimensions $d>2$. {\it This
is why the critical mass of isothermal spheres in $d=2$ shares some
analogies with the Chandrasekhar mass in $d=3$ since they both
correspond to critical polytropes with index $n=n_{3}$}
\cite{wd}. Comparing Eq. (\ref{i12}) with Eq. (\ref{up9}) we find that
for $d\rightarrow 2$ and $n=n_{3}\rightarrow +\infty$, we have the
limit
\begin{eqnarray}
\lim_{n_{3}\rightarrow +\infty} n_{3}\omega_{n_{3}}=4.
\label{up18}
\end{eqnarray}
This limit can also be obtained from Eq. (79) of \cite{lang}. With
this relation, we find that the critical mass and the critical
temperature in $d=2$ given by Eq. (\ref{i13}) are particular cases of
Eqs. (\ref{up15}) and (\ref{up16}).

Finally, for $d=1$ with $n>0$ (and for $d=2$ with $0<n<+\infty$), the
GSP system always relaxes towards a complete polytrope which is a global
minimum of free energy. Thus there is no critical dynamics for $d<2$
(and for $d=2$ with $n\neq +\infty$).

\subsection{Box confined polytropes}
\label{sec_box}

For systems confined within a box of radius $R$, we need to integrate
the Lane-Emden equation (\ref{up7}) until the normalized box radius
$\xi=\alpha$ with
\begin{eqnarray}
\alpha=R/r_0=\left\lbrack\frac{S_{d}G\rho_0^{1-1/n}}{K(n+1)}\right\rbrack^{1/2}R.
\label{box1}
\end{eqnarray}
It is useful to define a dimensionless control parameter (the
definition of this parameter has been slightly changed with respect
to our previous paper \cite{lang}):
\begin{eqnarray}
\eta=M\left \lbrack \frac{nS_{d}^{1/n}G}{K(1+n)}\right
\rbrack^{n/(n-1)}\frac{1}{R^{(d-2)(n-n_{3})/(n-1)}}.
\label{box2}
\end{eqnarray}
In terms of the polytropic temperature (\ref{up14}), it can be rewritten
\begin{equation}
\eta={G^{n/(n-1)}M\over \Theta^{n/(n-1)}R^{(d-2)(n-n_{3})/
(n-1)}}. \label{box3}
\end{equation}
Note that for $n\rightarrow +\infty$, we have $\Theta=K=k_{B}T/m$
and the definitions (\ref{i9}) and (\ref{box3}) coincide. Using the
conservation of mass or the Gauss theorem, we get \cite{lang}:
\begin{eqnarray}
\eta=-n^{{n}/({n-1})}\alpha^{(n+1)/(n-1)}\theta'(\alpha), \quad (\alpha<\xi_1).
\label{box4}
\end{eqnarray}
This equation relates the central density to the mass (at fixed
temperature and box radius). In fact, this relation is valid only
for {\it incomplete polytropes} whose density profile is arrested by
the box (i.e. $\rho(R)>0$). For $n\ge n_5$, this is always the case.
For $0<n<n_5$, using the identity
\begin{eqnarray}
\frac{\alpha}{\xi_{1}}=\frac{R}{R_{*}},
\label{box5}
\end{eqnarray}
the polytrope is confined by the box if $R_{*}\ge R$,
i.e. $\alpha\le \xi_{1}$. For $R_{*}<R$, i.e. $\alpha>\xi_{1}$, we
have {\it complete polytropes} whose density profile vanishes before
the wall. In that case, we need to integrate the
Lane-Emden equation until the natural polytropic radius $\xi=\xi_{1}$.
For $\alpha>
\xi_{1}$, the relation (\ref{box4}) is replaced by
\begin{eqnarray}
\eta=n^{n/(n-1)}\omega_{n} \left (\frac{R_{*}}{R}\right
)^{(d-2)(n-n_{3})/(n-1)}\ (\alpha>\xi_1), \label{box6}
\end{eqnarray}
which is equivalent to the mass-radius relation (\ref{up12}). Using
Eq. (\ref{box5}), it can be expressed in terms of $\alpha$, giving the
relation between the mass and the central density (at fixed
temperature) for complete polytropes.  Finally, the intermediate case
is $R_*=R$, i.e. $\alpha=\xi_{1}$, at which the density profile
vanishes precisely at the box radius. In that case, we have
\begin{eqnarray}
\eta=n^{n/(n-1)}\omega_{n}\quad (\alpha=\xi_1).
\label{box7}
\end{eqnarray}

\begin{figure}[htbp]
\centerline{
\includegraphics[width=8cm,angle=0]{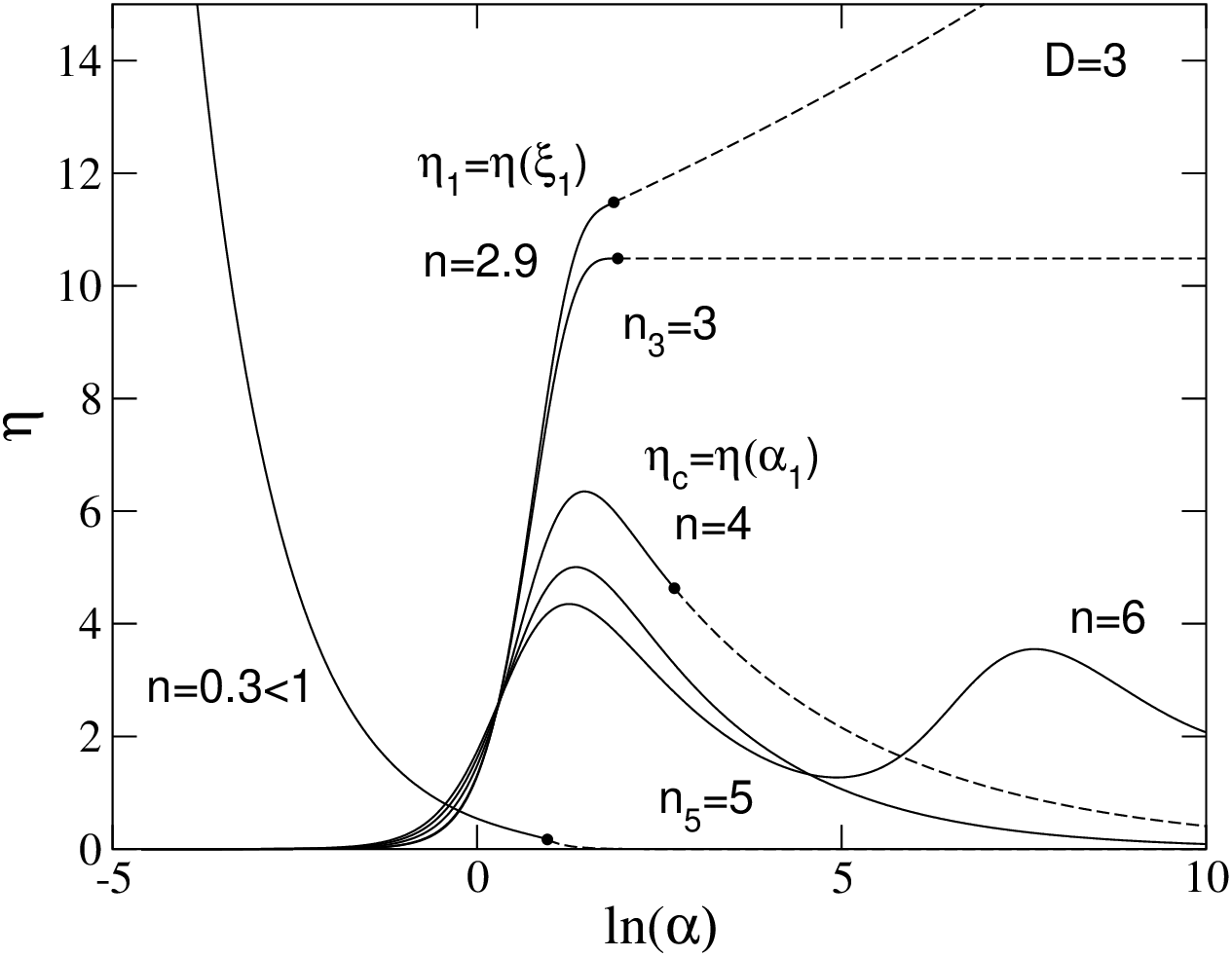}
} \caption[]{Series of equilibria for box-confined polytropes with
different index (the figure is done for $d=3$). The full lines
($\alpha<\xi_{1}$) correspond to incomplete polytropes whose profile is
arrested by the box and the dashed lines ($\alpha>\xi_{1}$) correspond
to complete polytropes that are self-confined.}
\label{alphaeD3}
\end{figure}

The relation $\eta(\alpha)$ defines the series of equilibria
containing incomplete (for $\alpha<\xi_{1}$) and complete (for
$\alpha>\xi_{1}$) polytropes. It gives the mass $M$ as a function of
the central density (for a fixed temperature $\Theta$ and box radius
$R$) or the temperature $\Theta$ as a function of the density contrast
(for a fixed mass $M$ and radius $R$).  Different examples of curves
$\eta(\alpha)$ are represented in Fig. \ref{alphaeD3} for various
indices in $d=3$:

$\bullet$ For $n<n_{3}$, the series of equilibria $\eta(\alpha)$ is
monotonic. Since polytropic spheres are stable in absence of gravity
(corresponding to $\alpha\rightarrow 0$) and since there is no turning
point, the Poincar\'e argument implies that all the polytropes are
stable. It can be shown furthermore that they are fully stable (global
minima of free energy at fixed mass) so that the GSP system will tend
to a steady state for $t\rightarrow +\infty$. For
$\eta<\eta_{1}=\eta(\xi_{1})=n^{n/(n-1)}\omega_{n}$, the GSP system tends to
an incomplete polytrope confined by the box.  For $\eta>\eta_{1}$, the
GSP tends to a complete polytrope with radius $R_{*}<R$. This has been
illustrated numerically in Fig. 21 of
\cite{lang} for $n=3/2$ in $d=3$. This index corresponds to a
classical white dwarf star in astrophysics. If we remove the box, the
GSP system always tends to the complete polytrope.

$\bullet$ For $n>n_{3}$, the series of equilibria $\eta(\alpha)$
presents a turning point at $\eta_{c}=\eta(\alpha_{1})$.  According to
the Poincar\'e turning point argument, configurations with
$\alpha>\alpha_{1}$ are unstable (saddle points of free energy at
fixed mass). This concerns in particular the case of complete
polytropes for $n_3<n<n_5$ (corresponding to $\alpha=\xi_{1}$), the
Schuster polytrope $n=n_{5}$ and the singular polytropic spheres for
$n\ge n_{5}$ (corresponding to $\alpha=+\infty$). Configurations with
$\alpha<\alpha_{1}$ are metastable (local minima of free energy at
fixed mass) and they exist only for $\eta\le\eta_c$. There is no
global minimum of free energy for $n>n_{3}$. For $\eta\le \eta_{c}$,
depending on the form of the initial density profile, the GSP system
can either relax towards an incomplete polytrope confined by the box
(metastable) or collapse.  For $\eta>\eta_{c}$, the GSP system undergoes
gravitational collapse. This self-similar collapse has been studied in
detail in \cite{lang}. It is very similar to the self-similar collapse
of isothermal systems in $d>2$ corresponding to $n\rightarrow
+\infty$.  If we remove the box, the GSP system can either collapse or
evaporate depending on the initial condition (this will be illustrated
numerically in Sec. \ref{sec_evaporation}).

$\bullet$ The case $n=n_{3}$ is critical and will be studied in detail in this
paper. For the critical index $n=n_{3}$, the control parameter is
independent on the box radius and can be written
\begin{eqnarray}
\eta= M\left ( \frac{G}{\Theta}\right )^{n_3/(n_3-1)}.
\label{box8}
\end{eqnarray}
In terms of the critical mass (\ref{up15}) or critical temperature
(\ref{up16}), we have
\begin{equation}
\eta=n_{3}^{n_3/(n_3-1)}\omega_{n_{3}}\frac{M}{M_{c}}=n_{3}^{n_3/(n_3-1)}\omega_{n_{3}}
\left (\frac{\Theta_{c}}{\Theta}\right )^{n_{3}/(n_{3}-1)}.
\label{box9}
\end{equation}
For incomplete polytropes with $\alpha<\xi_{1}$, the relation
$\eta(\alpha)$ between the central density, the mass and the
temperature is given by Eq. (\ref{box4}). Their density profile is
given by Eq. (\ref{up17}) where $r_0$ is determined by
$(\Theta_c/\Theta)^{d/2}=M/M_c=-(1/\omega_{n_3})(R/r_0)^{d-1}\theta'(R/r_0)$,
equivalent to relation (\ref{box4}), so the central density is now
determined by the mass $M$ or the temperature $\Theta$.  Complete
polytropes with $\alpha\ge
\xi_{1}$ exist for a unique value of the control parameter
\begin{eqnarray}
\eta_{c}=n_{3}^{n_3/(n_3-1)}\omega_{n_{3}}.
\label{box10}
\end{eqnarray}
This corresponds to the critical mass $M=M_c$ or critical temperature
$\Theta=\Theta_c$.  Equilibrium states exist only for $\eta\le
\eta_{c}$, i.e $M\le M_c$ or $\Theta\ge \Theta_c$. For $\eta<\eta_{c}$,
they are fully stable (global minima of free energy at fixed mass). In
that case, the GSP system relaxes towards an incomplete polytrope
confined by the box. For $\eta=\eta_{c}$, i.e $M=M_c$ or
$\Theta=\Theta_c$, we have an infinite family of steady states
parameterized by their central density $\alpha\ge\xi_1$ or equivalently
by their radius $R_{*}\le R$. They are marginally stable
($\delta^{2}F=0$). For $\eta>\eta_{c}$, i.e $M>M_c$ or
$\Theta<\Theta_c$, the GSP system undergoes gravitational
collapse. The collapse dynamics is expected to be similar to the
critical collapse of isothermal systems with $n\rightarrow +\infty$ in
$d=2$ (see below). If we remove the box, the solution of the GSP
system evaporates for $\eta<\eta_c$, i.e.  $M<M_{c}$ or
$\Theta>\Theta_c$ and collapses for $\eta>\eta_c$, i.e. for $M>M_{c}$
or $\Theta<\Theta_c$. These different regimes will be studied in
detail in Secs. \ref{sec_collapse} and
\ref{sec_evaporation}.

\begin{figure}[htbp]
\centerline{
\includegraphics[width=8cm,angle=0]{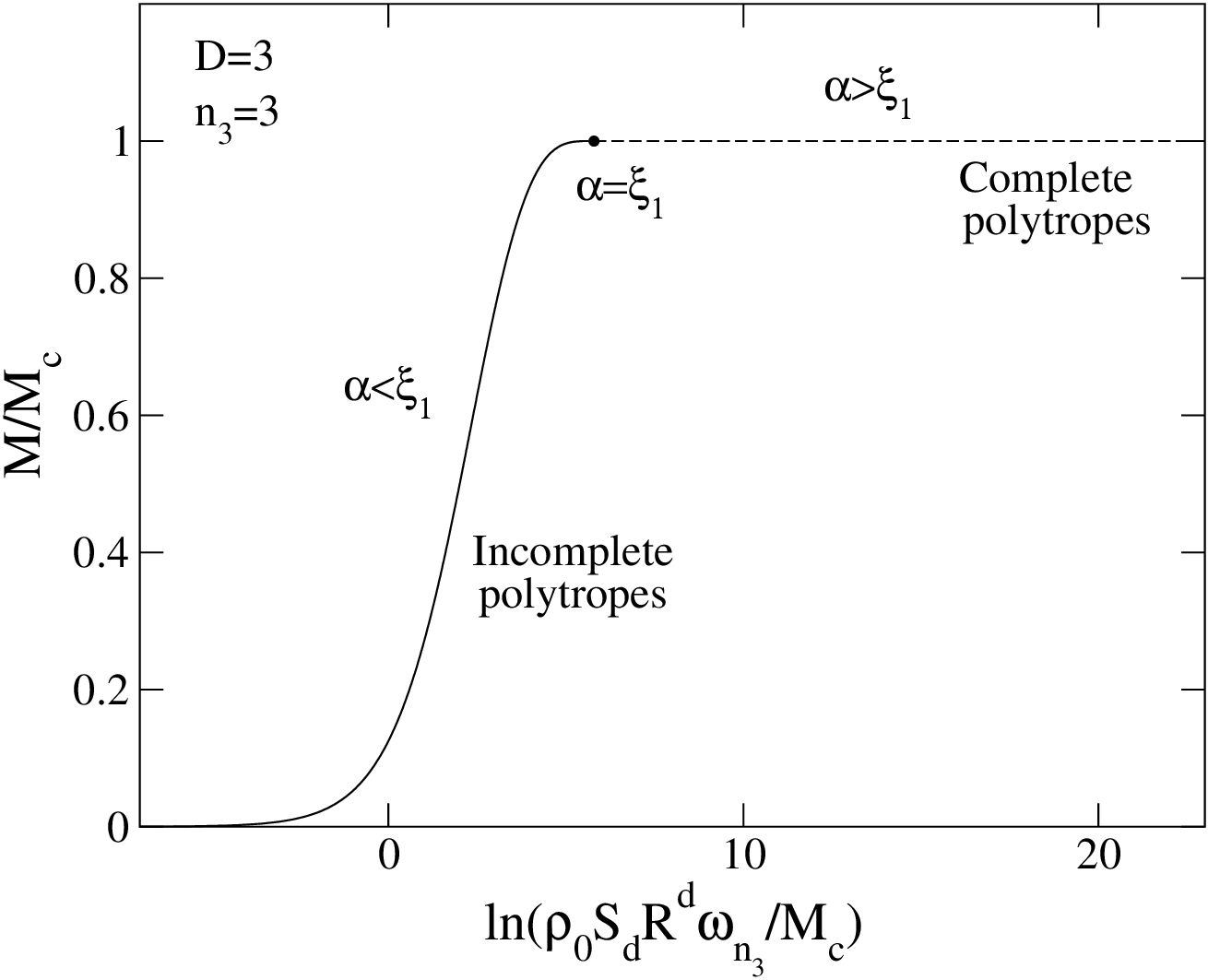}
} \caption[]{Mass as a function of the central density for
box-confined self-gravitating polytropic spheres with critical index
$n=n_{3}=3$ in $d=3$. Incomplete polytropes with $\rho(R)>0$ are
represented by a solid line and complete polytropes with $R_{*}\le R$
are represented by a dashed line. For $\rho_0\rightarrow +\infty$, the
density profile tends to a Dirac peak. Equilibrium states exist only
for $M\le M_c$. For $M>M_c$ the system undergoes gravitational
collapse.  The curve represents $\eta(\alpha)/\lbrack
n_3^{n_3/(n_3-1)}\omega_{n_3}\rbrack$ so it also gives the inverse
temperature $(\Theta_c/\Theta)^{n_3/(n_3-1)}$ as a function of the
density contrast ${\cal R}(\alpha)$ for a fixed mass. }
\label{alphaeta3}
\end{figure}

\begin{figure}[htbp]
\centerline{
\includegraphics[width=8cm,angle=0]{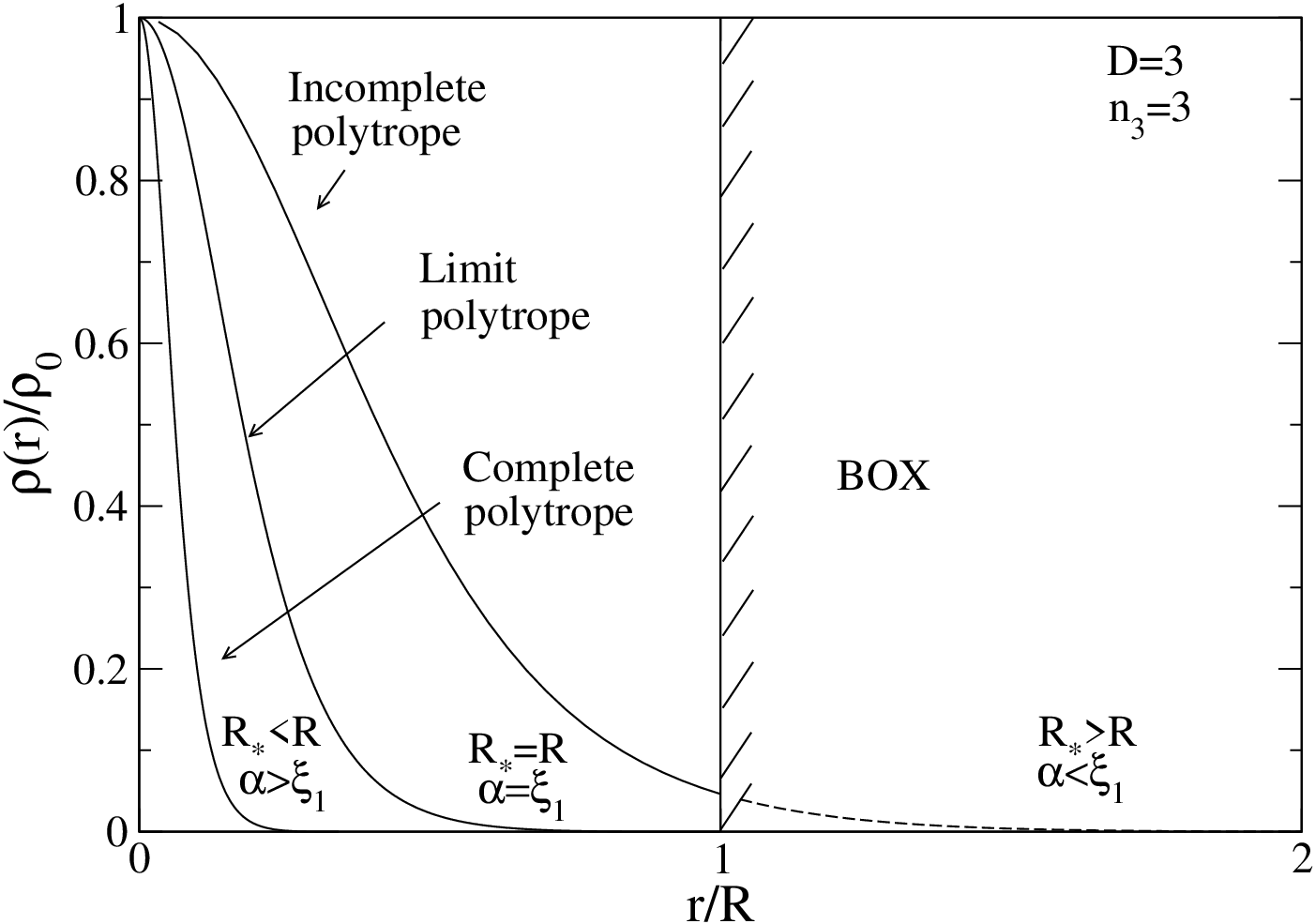}
} \caption[]{Density profiles of complete and incomplete polytropes
for the critical index $n_{3}=3$ in $d=3$. We have considered three
values of the central density
$\rho_{0}=(M_c/S_{d}R^d\omega_{n_3})\alpha^d$ corresponding to
$\alpha=3<\xi_{1}$ (incomplete polytrope: $R_{*}>R$, $M<M_{c}$),
$\alpha=\xi_{1}=6.89685...$ (limit polytrope: $R_{*}=R$, $M=M_{c}$),
and $\alpha=20>\xi_{1}$ (complete polytrope: $R_{*}<R$, $M=M_{c}$).}
\label{box}
\end{figure}

\begin{figure}[htbp]
\centerline{
\includegraphics[width=8cm,angle=0]{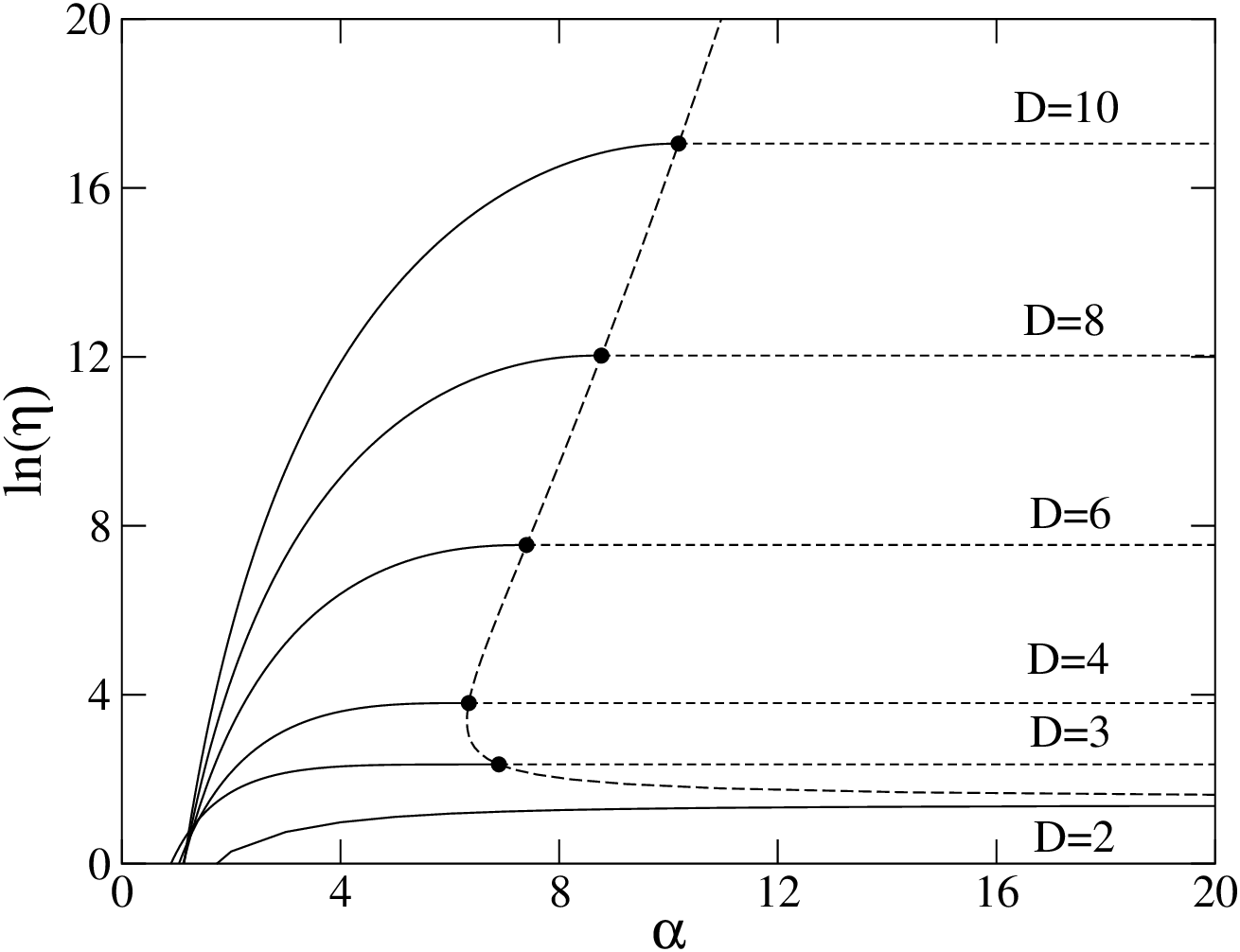}
} \caption[]{Mass as a function of the central density for critical
polytropes $n=n_3$ for different
dimensions of space. We have plotted
$\eta=n_{3}^{d/2}\omega_{n_{3}}M/M_c$ as a function of
$\alpha=(\rho_{0}S_{d}R^{d}\omega_{n_3}/M_c)^{1/d}$. The maximum mass
is reached at the bullet corresponding to $\alpha=\xi_{1}$,
$\eta=\eta_{c}=n_{3}^{d/2}\omega_{n_{3}}$.}
\label{etaalpha}
\end{figure}

The mass-central density relation (for a fixed temperature) of
box-confined self-gravitating polytropic spheres with critical index
$n=n_{3}$ is plotted in Fig. \ref{alphaeta3} and the corresponding
density profiles (illustrating the notion of complete and incomplete
polytropes) are plotted in Fig. \ref{box}. We note the striking
analogy with the mass-central density relation of white dwarf stars in
Fig. \ref{massedensiteD3}. Indeed, ultra-relativistic white dwarf
stars are equivalent to polytropes with critical index $n=n_{3}=3$ in
$d=3$. In this context, the critical mass $M_c$ corresponds to the
Chandrasekhar limit. We emphasize, however, that we are considering
here pure critical polytropes enclosed within a box while in
Sec. \ref{sec_wdp} we considered self-confined {\it partially}
relativistic white dwarf stars for which a box is not needed. It is
only when $M\rightarrow M_{Chandra}$ (ultra-relativistic limit) that
they become equivalent to pure polytropes. Furthermore, at
$M=M_{Chandra}$ for white dwarf stars, the only steady state is a
Dirac peak while at $M=M_c$ for pure critical polytropes, we have an
infinite family of steady states with different central densities (the
same difference holds between critical polytropes $n=n_{3}$ in $d>2$
and isothermal spheres $n=n_3=+\infty$ in $d=2$; compare Figs.
\ref{alphaeta3} and \ref{mrho}). Finally, in Fig. \ref{etaalpha}, 
we plot the mass as
a function of the central density for different dimensions of space
$d$. This figure illustrates in particular the connexion between the
critical mass in $d=3$ reached for a finite value of the central
density and the critical mass in $d=2$ reached for an infinite value
of the central density.

\section{The critical index from dimensional analysis}
\label{sec_dim}

It is instructive to understand the origin of the critical index
$\gamma_{4/3}=2(d-1)/2$ or $n_{3}=d/(d-2)$ from simple dimensional
analysis. Here we consider unconfined systems in $d$ dimensions with
arbitrary value of $\gamma$. The polytropic Smoluchowski-Poisson
system can be written
\begin{equation}
\frac{\partial\rho}{\partial t}=\nabla\cdot \left \lbrack
\frac{1}{\xi}\left (K\gamma\rho^{\gamma-1}\nabla
\rho+\rho\nabla\Phi\right )\right\rbrack \equiv -\nabla\cdot {\bf
J}, \label{d1}
\end{equation}
\begin{equation}
\Delta\Phi=S_{d}G\rho.
\label{d2}
\end{equation}
The current ${\bf J}={\bf J}_{d}+{\bf J}_{g}$ appearing in the
Smoluchowski equation is the sum of two terms: a  diffusion current ${\bf
J}_{d}=-K\gamma\rho^{\gamma-1}\nabla \rho$ and a gravitational
drift ${\bf J}_{g}=-\rho\nabla\Phi$. Based on dimensional analysis, the
diffusion current can be estimated by
\begin{equation}
J_{d}\sim + K\gamma
(M/L^d)^{\gamma-1}(\rho/L)\sim + (1/L)^{d(\gamma-1)+1},
\label{d3}
\end{equation}
and the drift term by
\begin{equation}
J_{g}=-\rho
GM/L^{d-1}\sim - (1/L)^{d-1},
\label{d4}
\end{equation}
where $M$ is the mass of the system and $L$ is the characteristic size
of the system.

The system will collapse to a point if gravity overcomes
(anomalous) diffusion, i.e. $|J_{g}|\gg |J_{d}|$, when $L\rightarrow
0$. This will be the case if $d-1>d(\gamma-1)+1$,
i.e. $\gamma<\gamma_{4/3}$. Conversely, if $\gamma>\gamma_{4/3}$, the
diffusion term can stabilize the system against gravitational collapse
so that the system can be in stable equilibrium.  The system will
evaporate to infinity if (anomalous) diffusion overcomes gravity,
i.e. $|J_{d}|\gg |J_{g}|$, when $L\rightarrow +\infty$. This will be
the case if $d(\gamma-1)+1<d-1$, i.e. if
$\gamma<\gamma_{4/3}$. Conversely, if $\gamma>\gamma_{4/3}$, the
gravitational attraction can prevent evaporation so that the system
can be in stable equilibrium. In conclusion, we find that the system
can be in a stable equilibrium state iff $\gamma>\gamma_{4/3}$,
i.e. $1/n>1/n_3$. In the opposite case, the system can either collapse
to a point or evaporate to infinity. By this very simple argument, we
recover the stability criterion of self-gravitating polytropic spheres
obtained by other methods (see Appendix B of \cite{wd}).

The critical case is obtained when $J_d\sim J_g$ implying
$d(\gamma-1)+1=d-1$, i.e. $\gamma=\gamma_{4/3}$ or, equivalently,
$n=n_{3}$. In that case, the stability of the system will depend on
its mass. The system will collapse to a point if gravity overcomes
diffusion, i.e. $|J_{g}|\gg |J_{d}|$, when $L\rightarrow 0$. This will
be the case if $M>M_{c}$, where $M_{c}\sim (K/G)^{d/2}$ is a critical
mass. The system will evaporate to infinity (in an unbounded domain)
if (anomalous) diffusion overcomes gravity, i.e. $|J_{d}|\gg |J_{g}|$,
when $L\rightarrow +\infty$. This will be the case if $M<M_{c}$.
Therefore, at the critical index $\gamma=\gamma_{4/3}$ i.e. $n=n_{3}$,
the system collapses if $M>M_c$ and evaporates if $M<M_c$. Again, this
is fully consistent with the results obtained in Appendix B of
\cite{wd}.

\section{Collapse dynamics}
\label{sec_collapse}

For $0<n<n_3$ in a space with $d\ge 2$ dimensions, the GSP system
tends to an equilibrium state. For $n\ge n_{3}$, it can undergo
gravitational collapse. For $n>n_{3}$ with $d>2$, the collapse is
self-similar as studied in \cite{lang} (the case of negative indices
$n<0$ is studied in \cite{logotropes}). In the present section, we consider
the collapse dynamics of self-gravitating Langevin particles
associated with the {\it critical} index $n_3=d/(d-2)$ in $d\ge 2$
dimensions which presents non trivial features.

\subsection{Generalities: self-similar analysis}
\label{sec_g}

From now on, we adopt normalized variables such that
$G=M=R=\xi=1$.  The unique control parameter is the temperature
$\Theta$. For spherically symmetric solutions, using the Gauss theorem, the GSP
system can be written in the form of an integrodifferential equation
\begin{eqnarray}
\frac{\partial\rho}{\partial t}=\frac{1}{r^{d-1}}
\frac{\partial}{\partial r}\biggl\lbrace r^{d-1}\biggl\lbrack
(S_{d}\rho)^{1/n}\Theta \frac{\partial\rho}{\partial r}\nonumber\\
+\frac{\rho}{r^{d-1}}\int_{0}^{r}\rho(r')S_{d}r'^{d-1}\,dr'
\biggr\rbrack\biggr\rbrace. \label{g1}
\end{eqnarray}
Introducing
the mass within a sphere of radius $r$
\begin{equation}
M(r,t)=\int_{0}^{r}\rho(r')S_{d}r'^{d-1}\,dr', \label{g2}
\end{equation}
the GSP system can be formulated through a unique
non-linear dynamical equation for $M(r,t)$:
\begin{eqnarray}
\frac{\partial M}{\partial t}=&&\Theta \biggl ({1\over
r^{d-1}}{\partial M\over\partial r}\biggr )^{1/n}\biggl\lbrack
{\partial^{2}M\over\partial r^{2}}-{d-1\over r}{\partial
M\over\partial r}\biggr\rbrack\nonumber
\\&& +{M\over r^{d-1}}{\partial M\over\partial r}. \label{g3}
\end{eqnarray}
If the system of total mass $M=1$ is confined within a box of radius
$R=1$, the appropriate boundary conditions are
\begin{equation}
M(0,t)=0,\qquad M(1,t)=1. \label{g4}
\end{equation}
If the system is not confined, the second condition should be replaced by
\begin{equation}
 M(\infty,t)=1. \label{g5}
\end{equation}
It is also convenient to introduce the function $s(r,t)=M(r,t)/r^{d}$
which has the same dimension as the density and which satisfies
\begin{equation}
{\partial s\over\partial t}=\Theta\biggl (r{\partial
s\over\partial r} +d s\biggr )^{1/n}\biggl
({\partial^{2}s\over\partial r^{2}}+{d+1\over r}{\partial
s\over\partial r}\biggr )+\biggl (r{\partial s\over\partial
r}+ds\biggr )s. \label{g6}
\end{equation}
For $n\rightarrow +\infty$, these equations reduce to those
studied in Refs. \cite{crs,sc} in the isothermal case.

When the system collapses, it is natural to look for self-similar solutions
of the form
\begin{equation}
\rho(r,t)=\rho_{0}(t)f\biggl ({r\over r_{0}(t)}\biggr ), \qquad
r_{0}=\biggl ({\Theta\over \rho_{0}^{1-1/n}}\biggr )^{1/2}.
\label{g7}
\end{equation}
The relation between the core radius $r_0$ and $\rho_0$
(proportional to the central density \footnote{The reader should be
aware that, in the sections dealing with the dynamics, $\rho_0$ and
$r_0$ do not exactly coincide with the quantities of the same name
introduced in the sections dealing with the statics (they usually
differ by a factor of proportionality).}) is obtained by requiring
that the diffusive term and the drift term in Eq. (\ref{g1}) scale
in the same way. This relation can be rewritten $\rho_0 r_0^{\alpha}\sim 1$
with
\begin{equation}
\alpha=\frac{2n}{n-1}.
\label{g8}
\end{equation}
In terms of the
mass profile, we have
\begin{equation}
M(r,t)=M_{0}(t)g\biggl ({r\over r_{0}(t)}\biggr ), \qquad {\rm
with}\qquad M_{0}(t)=\rho_{0}r_{0}^{d}, \label{g9}
\end{equation}
and
\begin{equation}
g(x)=\int_{0}^{x}f(x')S_{d}x'^{d-1}\,dx'. \label{g10}
\end{equation}
In terms of the function $s$, we have
\begin{equation}
s(r,t)=\rho_{0}(t)S\biggl ({r\over r_{0}(t)}\biggr ), \qquad {\rm
with}\qquad S(x)={g(x)\over x^{d}}. \label{g11}
\end{equation}
Inserting the ansatz (\ref{g11}) in Eq. (\ref{g6}) and using
Eq. (\ref{g7}), we obtain
\begin{equation}
\frac{1}{\rho_0^2}\frac{d\rho_0}{dt}=\alpha, \label{g12}
\end{equation}
and
\begin{equation}
\alpha S+xS'=(xS'+dS)^{1/n} \biggl (S''+{d+1\over x}S'\biggr
)+(xS'+dS)S. \label{g13}
\end{equation}
Assuming that Eq. (\ref{g13}) has a solution so that the
self-similar solution exists, Eq. (\ref{g12}) is readily integrated
in
\begin{equation}
\rho_0(t)=\frac{1}{\alpha}(t_{coll}-t)^{-1}, \label{g14}
\end{equation}
implying a finite time singularity. On the other hand, the invariant
profile has the asymptotic behavior $f(x)\sim x^{-\alpha}$ for
$x\rightarrow +\infty$.

\subsection{The two-dimensional isothermal case}
\label{sec_td}

In $d=2$ dimensions, the critical index is $n_{3}=+\infty$
corresponding to the isothermal case studied in \cite{sc} (in that case
$\Theta=T$). Since the study of the critical dynamics is rather
complicated, it can be useful to summarize our results, with some
complements and amplifications, before treating the case $d>2$.

In $d=2$, there exists a critical temperature $T_{c}=1/4$. If the
system is enclosed within a box and $T>T_c$, it relaxes to an
equilibrium distribution confined by the box. If the system is not
confined and $T>T_c$, an evaporation process develops which has been
studied in \cite{virial1}.  For $T=T_c$, the system undergoes
gravitational collapse. The evolution is self-similar and leads to a
Dirac peak containing the whole mass $M=1$ for $t\rightarrow
+\infty$. In a bounded domain, the central density grows exponentially
\cite{sc} rapidly with time and in an unbounded domain, the central
density increases logarithmically \cite{virial1} with time (and a tiny
fraction of mass is ejected at large distances to satisfy the moment
of inertia constraint at $T=T_{c}$). Note that the Dirac peak is also
the stationary solution of the SP system at $T=T_c$.

For $T<T_c$, and irrespectively of the presence of a confining box,
there is no steady state and the system collapses. Looking for an exact
self-similar solution of the form (\ref{g7}) we obtain $\rho_0 r_0^2=T$,
$\alpha=2=d$ and a scaling equation
\begin{equation}
\biggl (S''+{3\over x}S'\biggr
)+(xS'+2S)(S-1)=0. \label{td1}
\end{equation}
However, this equation does not have any physical solution for large
$x$. In fact, this could have been anticipated from the fact that the
scaling functions $s(x)$ and $f(x)$ should decay as $x^{-2}=x^{-d}$
for large $x$. Then, the total mass in the profile is of order $\rho_0
r_0^2\int^{1/r_0}x^{-2} x\,dx\sim \ln(1/r_0)$, which unphysically
diverges when $r_0$ goes to zero. Said differently, the scaling
profile at $t=t_{coll}$ is $\rho\propto r^{-2}$ so that the mass
$M=\int \rho(r)2\pi r dr$ diverges logarithmically for $r\rightarrow
0$. This logarithmic divergence is symptomatic of the formation of a
Dirac peak resulting from a pseudo self-similar collapse. In the
case $d=2$, this situation can be analyzed analytically in great
detail.

\begin{figure}[htbp]
\centerline{
\includegraphics[width=8cm,angle=0]{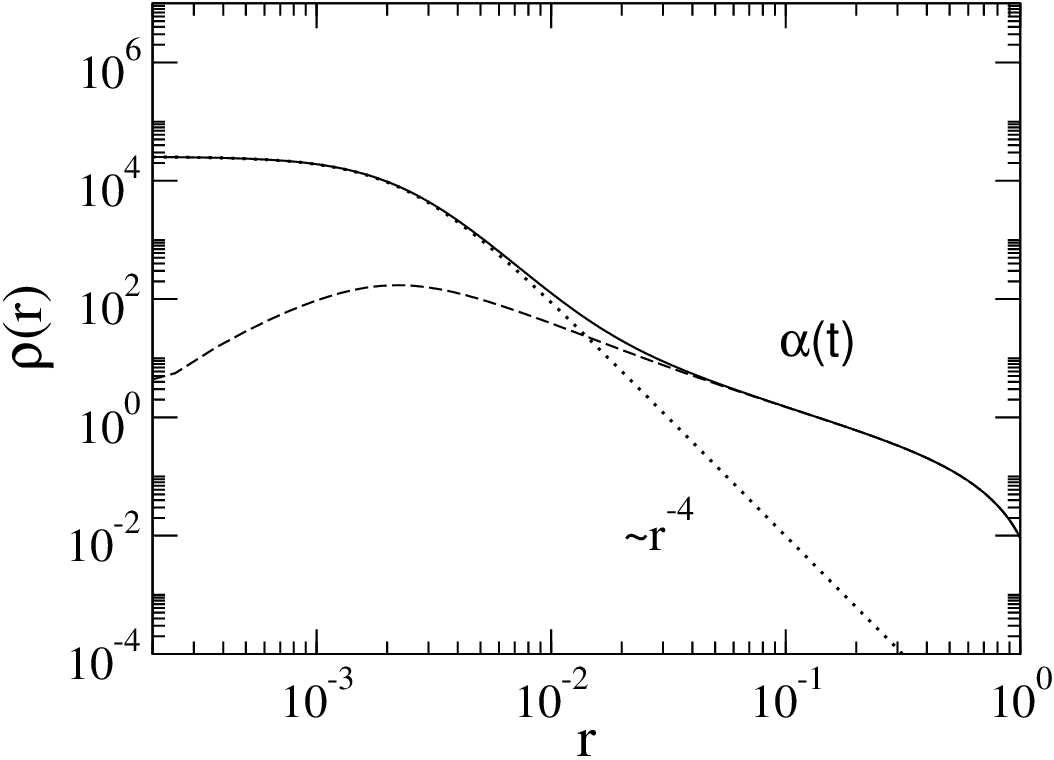} }
\caption[]{For $d=2$, $n=n_{3}=+\infty$, and deep into the collapse
regime for $T=T_c/2=1/8$, we plot the density profile (full line),
emphasizing its two components: the core is dominated by the
invariant scaling profile (dotted line) given analytically by
Eq.~(\ref{td3}) containing a mass $M_c=T/T_c$, and the halo obeys
pseudo-scaling  (dashed line) with an exponent $\alpha(t)$ tending
slowly to $d=2$ as $\rho_{0}\rightarrow +\infty$.} \label{D2SEUL}
\end{figure}

To that purpose, we note that the profile which cancels out the
r.h.s. of the SP system is exactly given by
\begin{equation}
M_1(r,t)=4T\frac{(r/r_0(t))^2}{1+(r/r_0(t))^{2}},
\label{td2}
\end{equation}
\begin{equation}
\rho_1(r,t)=\frac{4\rho_0(t)}{\pi}\frac{1}{(1+(r/r_0(t))^{2})^2},
\label{td3}
\end{equation}
with
\begin{equation}
\rho_{0}(t)r_{0}(t)^{2}=T.
\label{td4}
\end{equation}
If we consider time independent solutions ($\partial\rho/\partial
t=0$) and impose the conservation of mass, we recover the steady
solutions which exist for $T\ge T_{c}$ in a bounded domain (in that case
$r_0=(T/T_c-1)^{1/2}$) and for $T=T_c$ only in an infinite domain
(in that case we get a family of distributions parameterized by
$r_0$). However, in the present case, we consider the case $T<T_c$ and
seek the temporal evolution of $\rho_{0}(t)$ and $r_{0}(t)$. We argue
that the solution (\ref{td3}) gives the leading contribution of the density
profile in the core. This profile contains a mass $T/T_c$. We expect
that the collapse will lead to $\rho_{0}(t)\rightarrow +\infty$ and
$r_0(t)\rightarrow 0$ for $t\rightarrow t_{coll}$ (finite time
singularity). Then, we see that the profile (\ref{td3}) leads to a Dirac peak
with mass $T/T_c$, i.e.
\begin{equation}
\rho_1({\bf r},t)\rightarrow \frac{T}{T_{c}}\delta({\bf r}).
\label{td5}
\end{equation}
The excess of mass will be contained in the profile extending in the
halo. Therefore, we look for solutions of the form
\begin{eqnarray}
\label{td6}
\rho(r,t)&=&\rho_1(r,t)+\rho_2(r,t),\nonumber\\
&=&\rho_0(t) f_1(r/r_0(t))+\rho_0(t)^{\alpha(t)/2}f_2(r/r_0(t)).\nonumber\\
\end{eqnarray}
The first component has a scaling behavior and dominates in the center
of the collapse region. It leads to a Dirac peak containing a fraction
$M_c=T/T_{c}$ of the total mass $M=1$ at $t=t_{coll}$. The second component
obeys a pseudo-scaling and $f_2(x)\sim x^{-\alpha(t)}$ for large $x$,
with an effective scaling exponent $\alpha(t)$ which very slowly
approaches the value $2$ (expected from the naive self-similar analysis) when
 $t\rightarrow t_{coll}$. Thus, at $t=t_{coll}$, we get
\begin{equation}
\label{td7}\rho({\bf r},t)\rightarrow M_{c}\delta({\bf r})+\chi({\bf r},t),
\end{equation}
where $\chi(r)$ is singular at $r=0$ behaving roughly as $r^{-2}$.
In Fig. \ref{D2SEUL}, we illustrate this decomposition of the
density profile into two components.  It is shown in \cite{sc} that
the central density satisfies an equation of the form
\begin{equation}
\frac{1}{\rho_0}\frac{d\rho_0}{dt}\propto \rho_{0}^{\alpha(t)/2}, \label{td8}
\end{equation}
instead of Eq. (\ref{g12}), and that the effective scaling exponent
$\alpha(t)$ depends on the central density as
\begin{equation}
\epsilon(t) \equiv 1-\frac{\alpha(t)}{2}\sim \sqrt{\frac{\ln \ln
\rho_{0}(t)}{2\ln\rho_{0}(t)}}. \label{td9}
\end{equation}
This yields $\rho_{0}\sim (t_{coll}-t)^{-1+\epsilon(t)}$ or equivalently
\begin{equation}
\ln(\rho_{0}\tau)\sim -2\ln(r_{0}/\sqrt{\tau})\sim
\sqrt{\frac{|\ln\tau| \ln|\ln\tau|}{2}}, \label{td10}
\end{equation}
where we have noted $\tau=t_{coll}-t$.

Prior to our work \cite{sc}, and unknown to us at that time, Herrero
\& Velazquez
\cite{herrerobio} had investigated the same problem in the context of
chemotaxis using a different method based on match asymptotics. For
$T<T_c$ (as far as we know, they did not consider the case $T=T_c$
treated in \cite{sc}), they showed that the system forms a Dirac peak
of mass $M_c=T/T_c$ (within our notations) surrounded by a halo containing
the excess of mass. On a qualitative point of view, the two scenarii are
consistent. On a
quantitative point of view, however, the scaling laws
\begin{eqnarray}
\ln(\rho_{0}\tau)\sim -2\ln(r_{0}/\sqrt{\tau})\sim \sqrt{2|\ln\tau|}\nonumber\\
+\frac{1}{2}\left (1-\frac{1}{\sqrt{|\ln\tau|}}\right )\ln|\ln\tau|. \label{td11}
\end{eqnarray}
obtained by Herrero \& Velazquez (HV) are slightly  different from
ours (SC). They lead to an effective exponent given by
\begin{eqnarray}
1-\frac{\alpha(t)}{2}\sim \sqrt\frac{2}{\ln\rho_0}+\frac{1}{2}\left
(1-\frac{1}{\sqrt{\ln\rho_0}}\right )\frac{\ln
\ln\rho_0}{\ln\rho_0},\label{td12}
\end{eqnarray}
instead of Eq. (\ref{td9}). For the densities accessible numerically,
one gets $\alpha_{SC}(\rho_0=10^3)=1.252...$ while
$\alpha_{HV}(\rho_0=10^3)=0.751...$ and
$\alpha_{SC}(\rho_0=10^5)=1.348...$ while
$\alpha_{HV}(\rho_0=10^5)=1.017...$. Numerical simulations performed
in \cite{sc} show a good agreement with the predicted values of
$\alpha_{SC}$ for the densities accessible. However, in view of the
complexity of the problem, and of the logarithmic (and sub-logarithmic!)
corrections, it is difficult to understand the origin of the (slight)
discrepancy between the two approaches. In any case, they both show
that the collapse is not exactly self-similar but that the apparent
scaling exponent $\alpha(t)$ is a very slowly varying function of the
central density.

\subsection{The critical polytropic case with $d>2$}
\label{sec_cp}

\begin{figure}[htbp]
\centerline{ \includegraphics[width=8cm,angle=0]{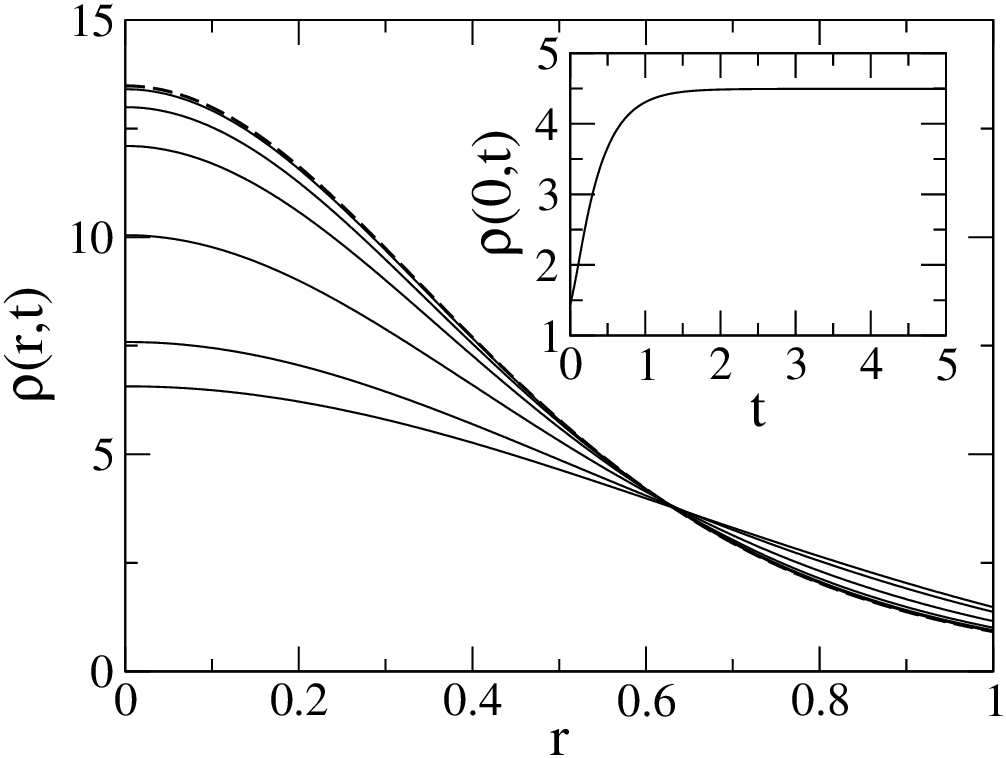} }
\caption[]{In $d=3$ and for $n_3=3$ and $\Theta=0.25>\Theta_c$ in a finite box
($R=1$), we show the density at successive times, illustrating the
convergence to the equilibrium density profile (dashed line). The
insert illustrates the exponentially fast saturation of the central
density for $\Theta>\Theta_c$, whereas a slower algebraic saturation is expected right at $\Theta=\Theta_c$.}
\label{coll1}
\end{figure}

We now consider the critical index $n=n_{3}=d/(d-2)$ with $d>2$. There
exists a critical temperature $\Theta_{c}=1/\lbrack
n_3\omega_{n_3}^{(n_3-1)/n_3}\rbrack$ (in $d=3$, we have
$\Theta_{c}=0.20872...$).  If the system is
confined within a box and $\Theta>\Theta_c$, it relaxes to an
incomplete polytrope. This is illustrated in Fig.~\ref{coll1}.  If the
system is not confined and $\Theta>\Theta_c$, an evaporation process
develops which will be studied in the next section. In the confined
case, when the generalized temperature $\Theta$ reaches the value
$\Theta_c$, the equilibrium density profile vanishes exactly at
$R=1$. For $\Theta<\Theta_c$, and irrespectively of the presence of a
confining box, the system collapses.

We can naively look for self-similar solutions of the form described
in Sec. \ref{sec_g}.  For $n=n_3$, we find $\alpha=d$,
$\rho_{0}r_{0}^d=\Theta^{d/2}$ and the scaling equation
\begin{equation}
S''+{d+1\over x}S'+(xS'+dS)^{2/d}(S-1)=0. \label{cp1}
\end{equation}
It happens that as in the case ($d=2$, $n_3=\infty$), this equation
does not have any physical solution for large $x$. Again, this could
have been anticipated from the fact that the scaling functions $s(x)$
and $f(x)$ should decay as $x^{-2n_{3}/(n_{3}-1)}=x^{-d}$, for large
$x$. Then, the total mass in the profile is of order
\begin{equation}
\rho_0
r_0^d\int^{1/r_0}x^{-d} {\times} x^{d-1}\,dx\sim \ln(1/r_0)
 \label{cp2}
\end{equation}
which unphysically diverges when $r_0$ goes to zero. Said
differently, the scaling profile at $t=t_{coll}$ is $\rho\propto
r^{-d}$ so that the mass $M=\int \rho(r)S_{d} r^{d-1} dr$ diverges
logarithmically for $r\rightarrow 0$ \footnote{More generally, for a
polytrope of index $n$ we have $\rho\propto r^{-\alpha}$ at
$t=t_{coll}$, with $\alpha=2n/(n-1)$ so that the self-similar
solution exists provided that $\alpha-d+1<1$ leading to $1/n<1/n_3$
(i.e. $n>n_3$ for $d>2$). This is precisely the range of indices for
which the complete polytropes are dynamically unstable
\cite{lang}.}.

\begin{figure}[htbp]
\centerline{
\includegraphics[width=8cm,angle=0]{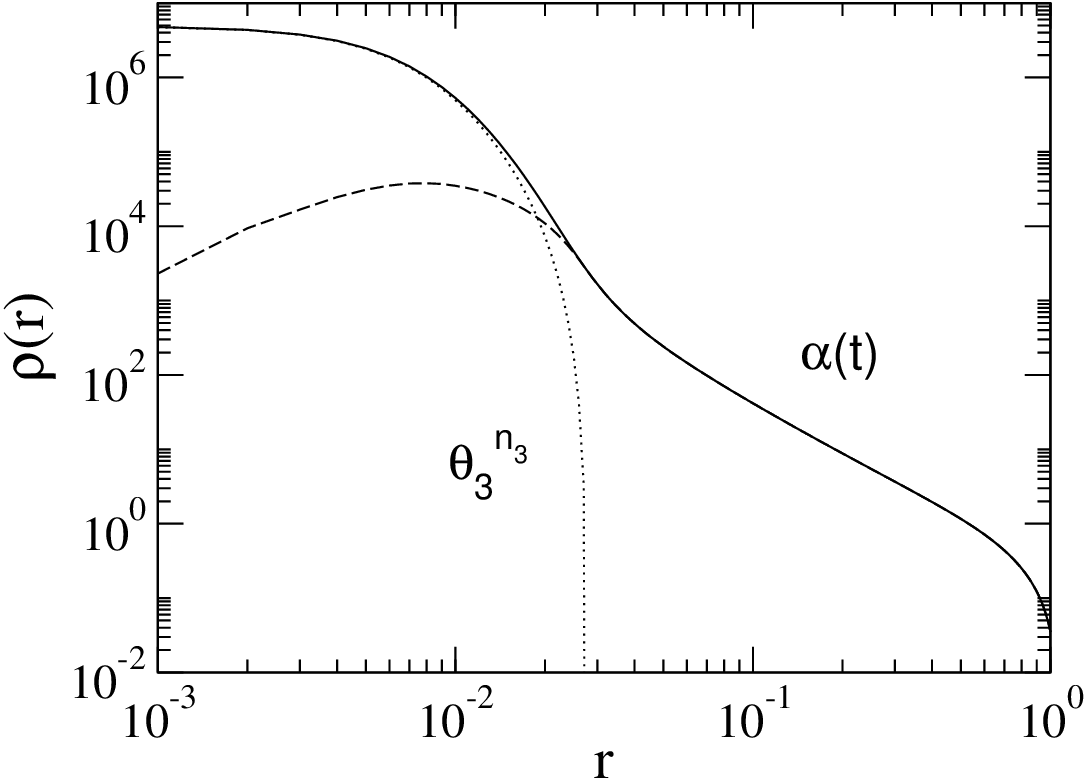} }
\caption[]{For $d=3$, $n=n_{3}=3$, and deep into the collapse regime
for $\Theta=0.75\Theta_c$, we plot the density profile (full line),
emphasizing its two components: the core is dominated by the bounded
invariant scaling profile (complete polytrope of index $n_3$)
containing a mass $M_c=(\Theta/\Theta_c)^{3/2}$ (dotted line), and
the halo obeys pseudo-scaling (dashed line) with an exponent
$\alpha(t)$ tending slowly to $d=3$ as $\rho_{0}\rightarrow
+\infty$.} \label{coll2}
\end{figure}

Hence, for $n=n_3$ in $d>2$, we expect a situation similar to what
was obtained for ($d=2$, $n_3=\infty$). However, the situation is
more difficult to analyze because the stationary profile is not
known analytically in the present case (this analytical profile was
at the basis of our analysis in \cite{sc}). Using the results of
Sec. \ref{sec_up}, the profile which cancels out the r.h.s. of the
GSP system is given by
\begin{equation}
\rho_1(r,t)=\frac{n_3^{d/2}}{S_{d}}\rho_0(t)\theta_{3}^{n_{3}}(r/r_{0}(t)),
\label{cp3}
\end{equation}
with
\begin{equation}
\rho_{0}(t)r_{0}(t)^{d}=\Theta^{d/2}.
\label{cp4}
\end{equation}
If we consider time independent solutions $(\partial\rho/\partial
t=0)$ and impose the conservation of mass, we recover the steady
solutions that exist for $\Theta\ge \Theta_{c}$ in a bounded domain
(in that case, we have
$(\Theta_c/\Theta)^{d/2}=-(1/\omega_{n_3})(R/r_0)^{d-1}\theta'(R/r_0)$)
and for $\Theta=\Theta_c$ only in an infinite domain (in that case we
get a family of distributions parameterized by $r_0$). However, in the
present case, we consider the case $\Theta<\Theta_c$ and seek the
temporal evolution of $\rho_{0}(t)$ and $r_{0}(t)$. We argue that the
solution (\ref{cp3}) gives the leading contribution of the density profile in
the core. This profile vanishes at $R_{*}(r)=\xi_{1}r_{0}(t)$,
has a central density $(n_3^{d/2}/S_{d})\rho_0(t)$ and contains a mass
(see Sec. \ref{sec_up}):
\begin{equation}
M_c=\left( \frac{\Theta}{\Theta_c}\right)^{d/2}.
\label{cp5}
\end{equation}
We expect that collapse will lead to $\rho_{0}(t)\rightarrow
+\infty$ and $r_{0}(t)\rightarrow 0$. Then, we see that the profile
(\ref{cp3}) tends to a Dirac peak with mass $M_c$, i.e.
\begin{equation}
\rho_1({\bf r},t)\rightarrow \left( \frac{\Theta}{\Theta_c}\right)^{d/2}\delta({\bf r}).
\label{cp6}
\end{equation}
The excess of mass will be contained in the profile extending in the
halo. Therefore, we look for solutions of the form
\begin{eqnarray}
\rho(r,t)&=&\rho_1(r,t)+\rho_2(r,t),\nonumber\\
&=&\rho_0(t) f_1(r/r_0(t))+\rho_0(t)^{\alpha(t)/d}f_2(r/r_0(t)).\nonumber\\
\label{cp7}
\end{eqnarray}
The first component \footnote{Defining $\rho_c(x)$ as the equilibrium
density profile at $\Theta=\Theta_c$, then the first component can be
written $\rho_1(r,t)=\frac{M_c}{r_0^d}\rho_c(r/r_0)$.} has a
scaling behavior and dominates in the center of the collapse
region. It leads to a Dirac peak containing a fraction
$M_c=(\Theta/\Theta_{c})^{d/2}$ of the total mass at $t=t_{coll}$. The
second component obeys a pseudo-scaling and $f_2(x)\sim
x^{-\alpha(t)}$ for large $x$, with an effective scaling exponent
$\alpha(t)$ which very slowly approaches the value $d$ (expected from
the naive self-similar analysis) when $t\rightarrow t_{coll}$. At
$t=t_{coll}$, the first component $\rho_1(r,t)$ tends to a Dirac peak
at the origin containing the mass $M_c$, whereas the second
component develops a singularity at $r=0$. Thus, we have
\begin{equation}
\label{cp8}\rho({\bf r},t)\rightarrow M_{c}\delta({\bf r})+\chi({\bf r},t),
\end{equation}
with $\chi(r)$ behaving roughly as $r^{-d}$.
In Fig.~\ref{coll2}, we
illustrate this decomposition of the density profile into two
components.
\begin{figure}[htbp]
\centerline{ \includegraphics[width=8cm,angle=0]{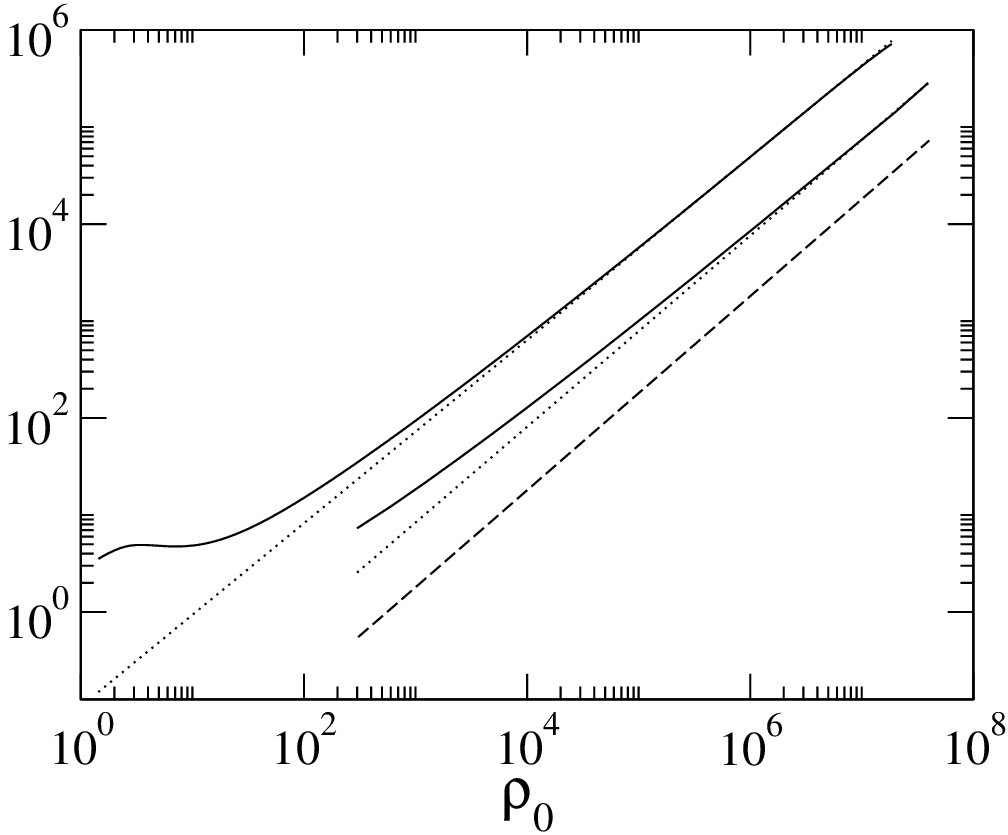}
} \caption[]{For $\Theta=0.75\Theta_c$ (here $d=3$, $n=n_{3}=3$), we
plot $\rho_0^{-1}(t)\frac{d\rho_0}{dt}$ (top full line) and
$\hat\rho_0(t)$ (bottom full line) as a function of $\rho_0(t)$.
Both grow with an effective exponent $\alpha/3\approx 0.93$ (dotted
lines), which slowly increases and should saturate to unity (the
dashed line has slope unity).} \label{coll3}
\end{figure}

In Fig.~\ref{coll3}, we show that perfect scaling which would imply
$\rho_0^{-1}(t)\frac{d\rho_0}{dt}\sim \rho_0$ is not obeyed.
Instead, in the accessible density range,
$\rho_0^{-1}(t)\frac{d\rho_0}{dt}$ decays with an apparent power-law
of $\rho_0$ which increases very slowly with time, but remains less
than unity. We expect to have a relation of the form
\begin{equation}
\frac{1}{\rho_0}\frac{d\rho_0}{dt}\propto \rho_{0}^{\alpha(t)/d}, \label{cp9}
\end{equation}
which is indeed confirmed by the numerics.  In Fig.~\ref{coll3}, we also plot
the central density in the pseudo-scaling component
\begin{equation}
\hat\rho_0(t)=\rho_0^{\alpha(t)/d}(t),
\label{cp10}
\end{equation}
which shows that the effective exponent $\alpha(t)$ slowly converges
to $\alpha=d$.

\begin{figure}[htbp]
\centerline{ \includegraphics[width=8cm,angle=0]{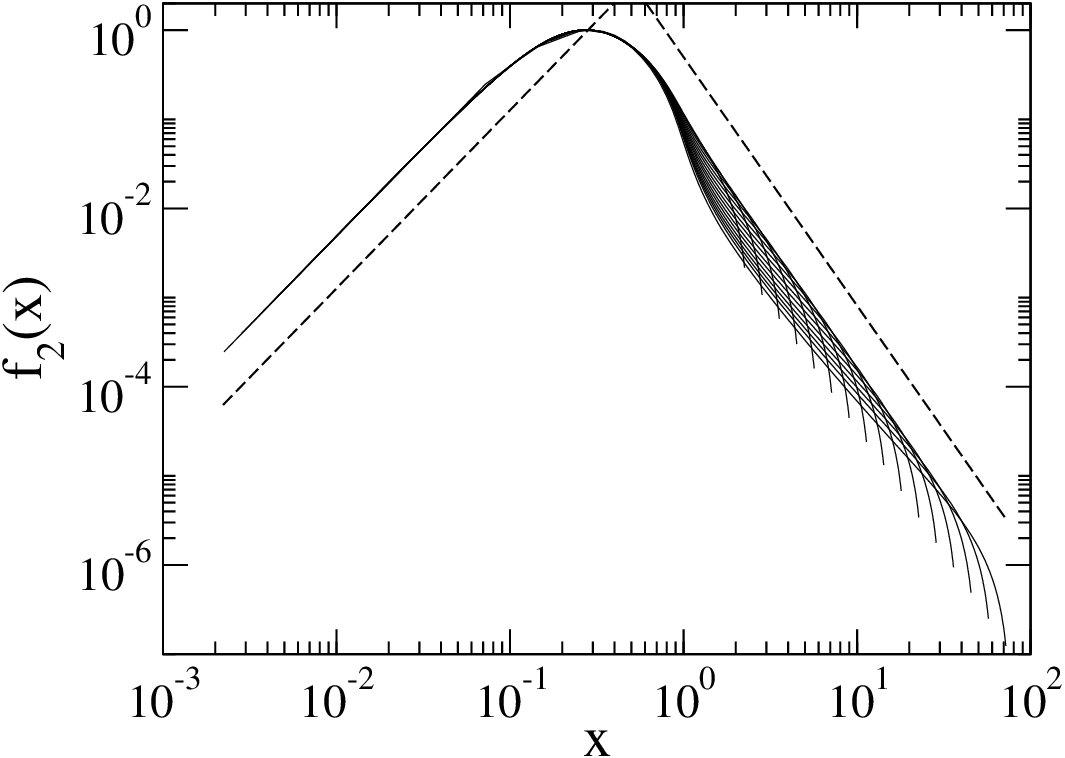}
} \caption[]{For $\Theta=0.75\Theta_c$ (here $d=3$, $n=n_{3}=3$), we
plot $f_2^{(\alpha)}(x)$ (as defined in the text) as a function of
$x=r/r_0$, for different times for which the central density evolves
from $10^2$ to $10^7$. Pseudo-scaling is observed. The envelop of
the tails decays with an apparent exponent $\alpha\approx 2.8$
(right dashed line), while the small $x$ behavior is quadratic (left
dashed line).} \label{coll4}
\end{figure}

Finally in Fig.~\ref{coll4}, we display the apparent scaling
behavior of $\rho_2(r,t)=\rho_0(t)^{\alpha(t)/d}f_2(r/r_0(t))$,
associated to a value of $\alpha\approx 2.8$, fully compatible with
the value obtained in Fig.~\ref{coll3} (in $d=3$).

\section{Evaporation dynamics in unbounded space}
\label{sec_evaporation}

\subsection{The case $n>n_3$}
\label{sec_sup}

\begin{figure}[htbp]
\centerline{ \includegraphics[width=8cm,angle=0]{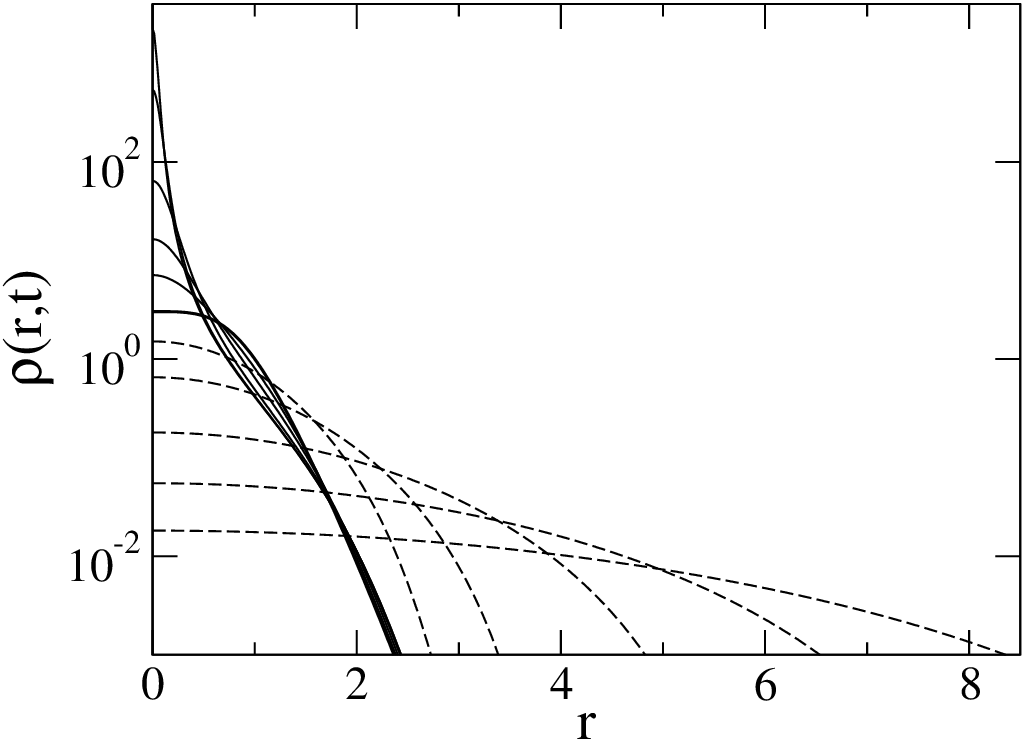}
} \caption[]{In $d=3$ and $n=5>n_3$, and for a given initial density
profile ($M(r)=r^3/({\rm e}^{-r^2}+r^2)^{3/2}$; fat line), we show
the collapse dynamics observed at $\Theta=0.15$ (full lines for
different times before $t_{coll}$) and the evaporation dynamics
observed at $\Theta=1$ (dashed lines for different times). For this
particular initial condition, we find $\Theta_*\approx 0.206$.}
\label{evap3}
\end{figure}

\begin{figure}[htbp]
\centerline{
\includegraphics[width=8cm,angle=0]{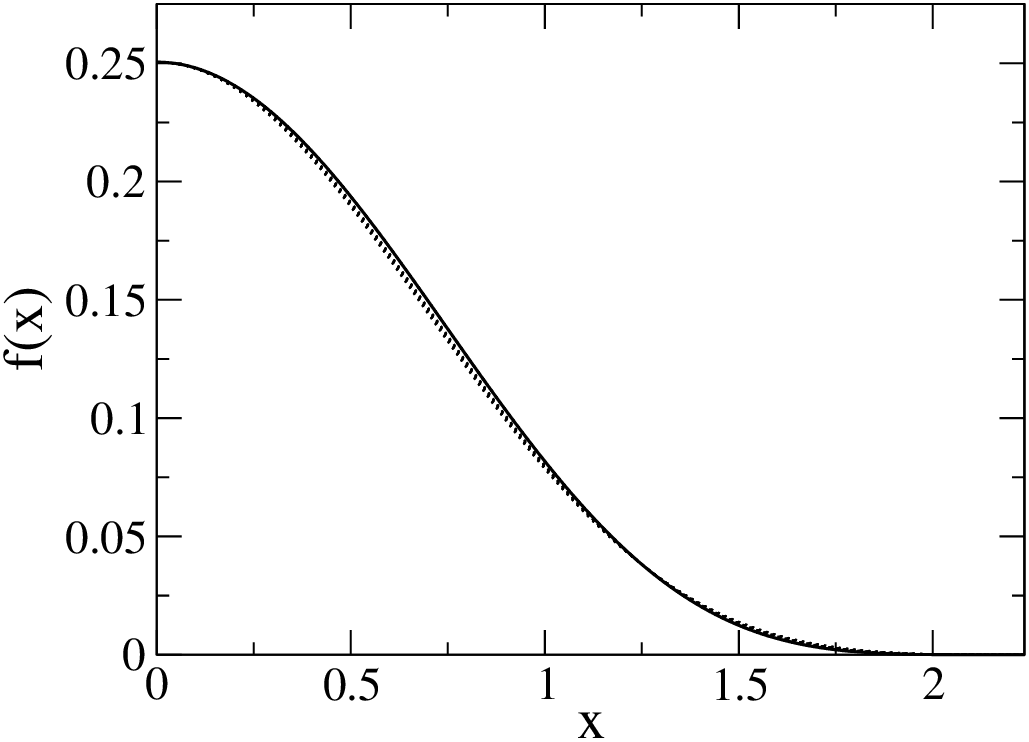} }
\caption[]{In $d=3$ and $n=5>n_3$, we present the evaporation
density data collapse at $\Theta=1$. As time proceeds, the effect of
gravity becomes negligible, and the scaling profiles converge to the
one corresponding to free diffusive evaporation (full line). This
is the Barenblatt solution whose invariant profile is a Tsallis
distribution of Eq.~(\ref{sup14}) with index $\gamma$.}
\label{evap4}
\end{figure}

When the system is not confined to a finite box, the nature of the
dynamics crucially depends on the value of the polytropic index $n$
with respect to $n_3$. As before, we consider $d\ge 2$ and $n>0$. If
$n<n_3$, there exists equilibrium solutions (fully stable complete
polytropes) which are reached for any initial density profile. If
$n>n_3$, depending on the initial density profile and on the
temperature, the system can collapse or evaporate. If $R_0$ is the
typical extension of the initial density profile containing a mass
$M$, one can form a quantity with the dimension of $\Theta$:
\begin{equation}
\label{sup1}
\Theta_*=\frac{G M^{(n-1)/n}}{R_0^{(d-2)(n-n_3)/n}},
\end{equation}
which plays the role of an effective critical temperature. If
$\Theta\ll\Theta_*$, the system should collapse as it would do if
confined in a box of typical radius $R_0$ \cite{lang}. If
$\Theta\gg\Theta_*$, the system should evaporate in the absence of an
actual confining box. Hence, for a given initial profile, there exists
a non universal $\Theta_*$ separating these two regimes.  We present
numerical simulations for the case $n>n_3$. In Fig.~\ref{evap3}, and
for a particular initial process, we illustrate the fact that
depending on the value of $\Theta$ with respect to a non universal
$\Theta_*$, the system can collapse or evaporate.  In the evaporation
regime and for $n>n_3$, a scaling analysis shows that gravity becomes
gradually irrelevant and that this process becomes exclusively
controlled by free (anomalous) diffusion. This fact is illustrated in
Fig.~\ref{evap4}.  Indeed, when the evaporation length
$r_{0}(t)\rightarrow +\infty$, we see from Eq. (\ref{g3}) that the
gravitational term becomes negligible in front of the diffusion term:
\begin{eqnarray}
{M\over r^{d-1}}{\partial M\over\partial r}\ll \Theta \biggl ({1\over
r^{d-1}}{\partial M\over\partial r}\biggr )^{1/n}
{\partial^{2}M\over\partial r^{2}}, \label{sup2}
\end{eqnarray}
if $d>d/n+2$, i.e. $n>n_3$. Therefore, for $t\gg 1$, the GSP system
reduces to the pure anomalous diffusion equation
\begin{equation}
\frac{\partial\rho}{\partial t}\simeq
\frac{K}{r^{d-1}}\frac{\partial}{\partial r}\left
(r^{d-1}\frac{\partial \rho^{\gamma}}{\partial r}\right
),\label{sup3}
\end{equation}
with $K=S_{d}^{\gamma-1}\Theta/\gamma$. This equation has self-similar
solutions that were first discovered by Barenblatt  \cite{barenblatt}
in the context of porous media. These solutions are closely related to
the form of generalized thermodynamics introduced by Tsallis \cite{tsallis}.

Using the original idea of Plastino \& Plastino \cite{pp}, we look for
solution of Eq. (\ref{sup3}) in the form of a Tsallis distribution
with index $\gamma$ and time dependent coefficients
\begin{equation}
\rho(r,t)=\frac{1}{Z}\rho_0(t)\left\lbrack
1-(\gamma-1)(r/r_0(t))^2\right\rbrack_{+}^{1/(\gamma-1)}.\label{sup4}
\end{equation}
For $\gamma>1$, i.e. $n>0$, we have a profile with compact support
where the density vanishes at
$r_{max}(t)=r_{0}(t)/\sqrt{\gamma-1}$. For $\gamma<1$, i.e. $n<0$, the
density decreases like $\rho\sim r^{-2/(1-\gamma)}$ and the total mass
is finite provided that $\gamma>\gamma_{1/3}\equiv (d-2)/d$,
i.e. $n<-d/2$. Requiring that the profile (\ref{sup4}) contains all the mass $M=1$,
and imposing
\begin{equation}
\rho_{0}(t)r_{0}(t)^d=1,\label{sup5}
\end{equation}
with find the normalization factor
\begin{equation}
Z\equiv \int_{0}^{+\infty}\lbrack
1-(\gamma-1)x^2\rbrack_{+}^{1/(\gamma-1)}S_{d}x^{d-1}dx.\label{sup6}
\end{equation}
Then, substituting the ansatz (\ref{sup4}) with Eq. (\ref{sup5})  in
Eq. (\ref{sup3}), we obtain
\begin{equation}
\dot \rho_0=-2dS_{d}^{\gamma-1}\Theta Z^{1-\gamma}\rho_0^{\gamma+2/d}.\label{sup7}
\end{equation}
Solving this equation with the initial condition $\rho({\bf
r},t=0)=\delta({\bf r})$, we get
\begin{equation}
\rho_0(t)=\frac{1}{\lbrack
2d(\gamma-\gamma_{1/3})S_{d}^{\gamma-1}\Theta  Z^{1-\gamma}
t\rbrack^{1/(\gamma-\gamma_{1/3})}}.\label{sup8}
\end{equation}
This is valid for $\gamma>\gamma_{1/3}$, i.e. $n>0$ or $n<-d/2$. We
note the scaling laws for large times:
\begin{equation}
\rho_0(t)\sim t^{-dn/(d+2n)}, \qquad r_0(t)\sim t^{n/(d+2n)}. \label{sup9}
\end{equation}

It is instructive to re-derive this solution in a different manner,
without pre-supposing the form of the solution. We look for general
self-similar solutions of the form
\begin{equation}
\rho(r,t)=\rho_{0}(t)f(r/r_{0}(t)). \label{sup10}
\end{equation}
We require that all the mass is in the profile (\ref{sup10}) and
impose the relation (\ref{sup5}), implying that
\begin{equation}
\int_{0}^{+\infty}f(x)S_{d}x^{d-1}\,dx=1.\label{sup11}
\end{equation}
Substituting the ansatz (\ref{sup10})  with Eq. (\ref{sup5}) in Eq.
(\ref{sup3}), and imposing the condition (\ref{sup7}) where $Z$ is
for the moment an arbitrary constant, we obtain the differential
equation
\begin{equation}
\frac{1}{x^{d-1}}\frac{d}{dx}\left
(x^{d-1}f^{\gamma-1}\frac{df}{dx}\right
)=-2Z^{1-\gamma}(xf'+df).\label{sup12}
\end{equation}
Noting the identity $x^{d-1}(xf'+df)=(x^d f)'$, this equation can be integrated into
\begin{equation}
f^{\gamma-2}\frac{df}{dx}+2Z^{1-\gamma}x=0. \label{sup13}
\end{equation}
This first order differential equation can again be readily
integrated. We can choose the constant of integration so as to
obtain a solution of the form
\begin{equation}
f(x)=\frac{1}{Z}\left\lbrack 1-(\gamma-1)x^2\right\rbrack_{+}^{1/(\gamma-1)}.\label{sup14}
\end{equation}
Finally, the normalization condition (\ref{sup11}) implies that $Z$
is given by Eq. (\ref{sup6}). It is interesting to realize that the
$q$-exponential function $e_{q}(x)=\lbrack
1+(q-1)x\rbrack_{+}^{1/(q-1)}$ introduced in the context of Tsallis
generalized thermodynamics stems from the simple differential
equation (\ref{sup13}) related to the anomalous diffusion equation
(\ref{sup3}). Indeed, the scaling solution of this equation can be
written
\begin{equation}
f(x)=\frac{1}{Z}e_{\gamma}(-x^2),\label{sup15}
\end{equation}
which generalizes the gaussian distribution obtained for the
ordinary diffusion equation recovered for $\gamma=1$.
The moments $\langle r^k\rangle$ of the distribution (\ref{sup10}) are
given by
\begin{equation}
\langle r^k\rangle(t)= r_0(t)^k\int_{0}^{+\infty}f(x)x^{k+d-1}S_d \,dx.\label{sup16}
\end{equation}
They exist provided that $k>-d$ for $\gamma\ge 1$ and provided that
$-d<k<2/(1-\gamma)-d$ for $\gamma<1$. They scale like $\langle
r^k\rangle\propto r_0^k\propto t^{nk/(d+2n)}$.

The Tsallis entropy is finite for $\gamma>\gamma_{3/5}=d/(d+2)$ and it scales like
\begin{equation}
S(t)-nM=-n\rho_0^{1/n}\int_0^{+\infty}f(x)^{\gamma}S_d
x^{d-1}dx\propto t^{-d/(d+2n)}. \label{sup17}
\end{equation}
On the other hand, for $d>2$, the potential energy $W=-1/(2S_d) \int
(\nabla\Phi)^2 d{\bf r}$ scales like
\begin{equation}
W\propto  \int_{0}^{+\infty} \left \lbrack
\frac{M(r)}{r^{d-1}}\right \rbrack^{2}r^{d-1}dr\propto
\frac{1}{r_{0}^{d-2}}\propto t^{-n(d-2)/(d+2n)}.\label{sup18}
\end{equation}
Comparing Eqs. (\ref{sup17}) and (\ref{sup18}), we see that the
potential energy is always negligible with respect to the entropy
for  $n>n_{3}$. Therefore, the Tsallis free energy behaves like
\begin{equation}
F(t)+nKM\propto t^{-d/(d+2n)},\label{sup19}
\end{equation}
for $t\rightarrow +\infty$. Note that for $n_3<n<+\infty$, the free
energy tends to a finite value $-nKM$ as the system spreads to
infinity. Alternatively, for the isothermal case $n= +\infty$, the free energy
is given by Eq. (95) of \cite{virial1} and it tends to $-\infty$.

We can use the identity (\ref{m4}) to derive the first correction in
the evolution of the moments $\langle r^k\rangle$ due to
self-gravity. To that purpose, we introduce the zeroth order
solution (\ref{sup4}) in the equation
\begin{eqnarray}
\label{sup20}
\frac{d\langle r^k\rangle}{dt}=k(k+d-2)\int  P r^{k-2} \,d{\bf r}\nonumber\\
-k \int_{0}^{+\infty} r^{k-d}M(r)\frac{\partial M}{\partial r}\,dr.
\end{eqnarray}
The first term gives, after integration, the pure anomalous scaling
\begin{eqnarray}
\label{sup21}
\langle r^{k}\rangle_{0} \propto t^{nk/(d+2n)}.
\end{eqnarray}
The second term gives, after integration, the first correction due
to gravity. If we write $\Delta  \langle r^{k}\rangle =\langle
r^{k}\rangle -\langle r^{k}\rangle_{0}$, we get
\begin{eqnarray}
\label{sup22}
\Delta \langle r^{k}\rangle \propto t^{\frac{n(k-d)}{d+2n}+1}.
\end{eqnarray}
Let us consider some particular cases: (i) for $n\rightarrow +\infty$,
we obtain $\Delta \langle r^{k}\rangle
\propto t^{(k-d)/2+1}$. If we  furthermore  consider the second moment $k=2$
(moment of inertia), we recover the scaling $\Delta \langle
r^{2}\rangle \propto t^{2-{d}/{2}}$ of \cite{virial1}. (ii) for $k=d$,
we find that $\Delta \langle r^{d}\rangle=-(d/2)t\propto t$ whatever
the index $n$ and the dimension of space $d$. (iii) For $n=n_3$,
gravitational effects are of the same order as diffusive effects and
$\langle r^k\rangle_{0}\propto \Delta \langle r^k\rangle\propto
t^{k/d}$. This case will be studied in detail in the next
section. (iv) Finally, let us introduce the number $k_0\equiv
d-d/n-2$. For $k<k_0$, $\Delta\langle r^k\rangle\rightarrow 0$; for
$k=k_0$, $\Delta\langle r^k\rangle\propto 1/t$; for $k>k_0$,
$\Delta\langle r^k\rangle\rightarrow +\infty$.

\subsection{The critical case $n=n_3$}
\label{sec_crit}

\begin{figure}[htbp]
\centerline{
\includegraphics[width=8cm,angle=0]{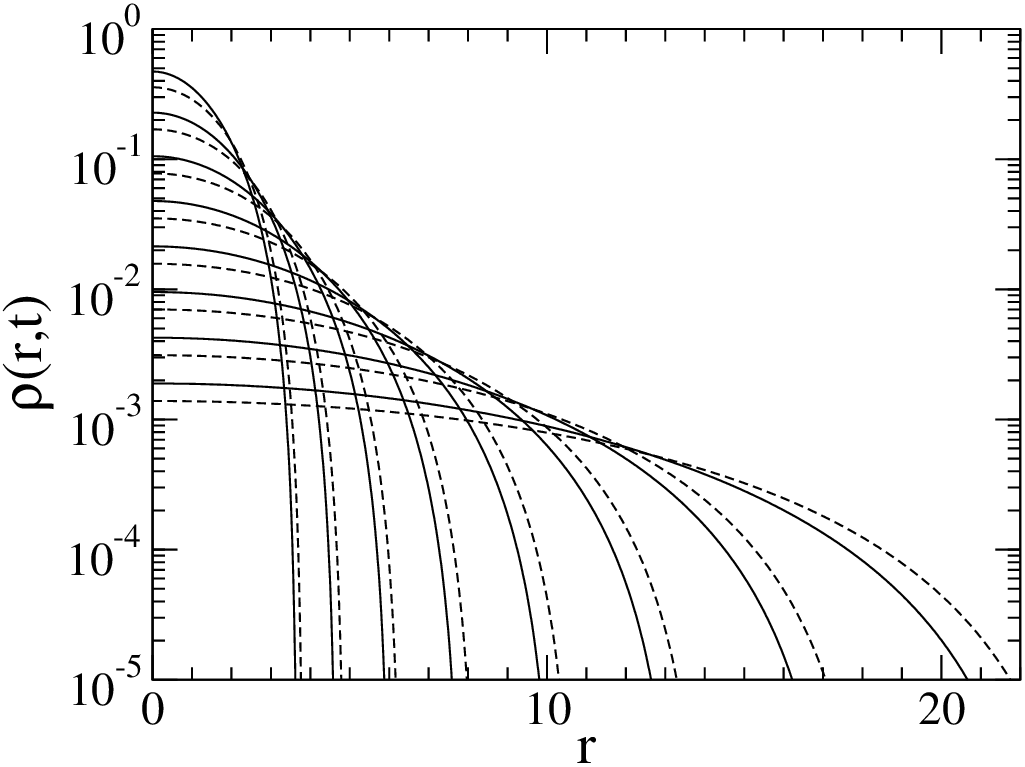} }
\caption[]{In $d=3$ and for $n=n_3=3$, we compare the evaporation
profiles at different times for $\Theta=1>\Theta_c$, for
self-gravitating particles (full lines), to the faster evaporation
dynamics when gravity is switched off (dashed lines). }
\label{evap1}
\end{figure}

Finally, for $n=n_3$, and since a critical $\Theta_c$ exists
irrespectively of the presence of a confining box, the system
collapses for $\Theta<\Theta_c$ and evaporates for
$\Theta>\Theta_c$. In the latter regime, gravity remains relevant
and evaporation is controlled by both gravity and the diffusion
process (see Fig.~\ref{evap1}). Mathematically, this arises from the
fact that there is an evaporation scaling solution for which all the
terms of Eq.~(\ref{g1}) scale in the same way. More specifically, we
expect an evaporation density profile of the form
\begin{equation}
\rho(r,t)=\rho_{0}(t)f\biggl ({r\over r_{0}(t)}\biggr ), \qquad
\rho_{0}(t)r_{0}(t)^d=1.
\label{crit1}
\end{equation}
The relation between the evaporation length $r_0$ and $\rho_0$
(proportional to the central density) is obtained by requiring that
the diffusive term and the drift term in Eq. (\ref{g1}) scale the same. In
terms of the mass profile, we have
\begin{equation}
M(r,t)= g\biggl ({r\over r_{0}(t)}\biggr )\quad {\rm with}\quad
g(x)=\int_{0}^{x}f(x')S_{d}x'^{d-1}\,dx',
\label{crit2}
\end{equation}
and in terms of the function $s$, we have
\begin{equation}
s(r,t)=\rho_{0}(t)S\biggl ({r\over r_{0}(t)}\biggr ), \qquad {\rm
with}\qquad S(x)={g(x)\over x^{d}}. \label{crit3}
\end{equation}
We require that all the mass is contained in the self-similar
profile \footnote{Looking for a self-similar solution of the form
(\ref{g7}) for any index $n$ and requiring that the diffusion and
gravity scale the same, we find that  $\rho_0 r_0^{\alpha}\sim 1$
where $\alpha$ is given by Eq. (\ref{g8}). The profile will contain
all the mass provided that $\rho_0 r_0^d\sim 1$ which implies
$\alpha=d$ leading to $n=n_{3}$. Thus, it is only for the critical
index that we can have a self-similar solution where the diffusion
and gravity scale the same and which contains all the mass.}, which
implies that
\begin{equation}
g(+\infty)=\int_{0}^{+\infty}f(x)S_{d}x^{d-1}\,dx=1.
\label{crit4}
\end{equation}
Inserting the ansatz (\ref{crit3}) in Eq. (\ref{g6}), using Eq.
(\ref{crit1}), and imposing
\begin{equation}
\frac{1}{\rho_{0}^{2}}\frac{d\rho_{0}}{dt}=-d\Theta, \quad {\rm
i.e.}\quad r_{0}^{d-1}\frac{dr_{0}}{dt}=\Theta,\label{crit5}
\end{equation}
we obtain the scaling equation (note the change of sign
compared to Eq.~(\ref{g13})) \footnote{{The scaling
equations (\ref{g13}) and (\ref{crit6}) have a very different
mathematical structure.  The scaling equation for collapse
(\ref{g13}), valid for $n>n_3$, leads to an eigenvalue problem for $S(x)$
\cite{sc,lang}. Indeed, it admits a physical solution for a unique
value of $S(0)$ equal to $S_{*}$ (say). For $S(0)<S_{*}$, the solution
becomes negative at some point, and for $S(0)>S_*$,
it diverges at a finite $x_0$. By contrast,
the scaling equation for evaporation (\ref{crit6}), valid for $n=n_3$,
admits a one parameter family of solutions parameterized by
$S(0)$. Then, the suitable value $S_{*}$ is selected by the
normalization condition (\ref{crit4}).}}:
\begin{equation}
S''+{d+1\over x}S'+(xS'+dS)^{2/d}\left (\frac{1}{\Theta}S+1\right )=0. \label{crit6}
\end{equation}
The evaporation radius is given by
\begin{equation}
r_0(t)=(d\Theta t)^{1/d}.\label{crit7}
\end{equation}
The moments scale like $\langle r^k\rangle\propto r_0^k\propto
(d\Theta t)^{k/d}$ and the free energy scales like $F(t)+n_3KM\propto
t^{-(d-2)/d}$.

If we consider the large temperature limit $\Theta\gg 1$ where the
diffusion term dominates on the gravitational drift, the foregoing
differential equation reduces to
\begin{equation}
S''+{d+1\over x}S'+(xS'+dS)^{2/d}=0. \label{crit8}
\end{equation}
In terms of the function $f$ it can be written
\begin{equation}
f^{-2/d} f'+\frac{x}{S_d^{(d-2)/d}}=0,\label{crit9}
\end{equation}
which is consistent with Eq.~(\ref{sup13}) up to the changes of
notations in Eqs. (\ref{sup7}) and (\ref{crit5}). We can either solve
this equation and impose the normalization condition (\ref{crit4}) or
make simple transformations in order to directly use the results of
Sec. \ref{sec_sup}. Indeed, let us set $\rho_{0}=\sigma\rho_{*}$ and
$r_{0}=\mu r_{*}$. We impose $\rho_* r_{*}^d=1$ leading to
$\sigma\mu^{d}=1$. On the other hand, we choose
$\sigma=2(S_{d}/Z)^{(d-2)/d}$ where $Z$ is defined by Eq. (\ref{sup6})
so that
$\dot{\rho}_{*}=-2d(S_{d}/Z)^{(d-2)/d}\Theta\rho_{*}^{2}$. Then,
$\rho=\rho_{*}f_{*}(r/r_{*})$ with $f_*(x)=\sigma f(x/\mu)$. Now,
$\rho_*$, $r_*$ and $f_*$ have been defined so as to coincide with the
functions $\rho_0$, $r_0$ and $f$ of Sec. \ref{sec_sup}. Thus, we get
$f(x)=(1/\sigma)f_*(\mu x)$ where $f_*$ is the function
(\ref{sup14}). Therefore, the normalized solution of Eq. (\ref{crit9})
with the present notations can be written
\begin{equation}
f(x)=\frac{1}{\sigma Z}\left\lbrack 1-\frac{d-2}{d}\mu^{2} x^2\right
\rbrack_{+}^{d/(d-2)},\label{crit10}
\end{equation}
with
\begin{equation}
\sigma \mu^{d}=1, \qquad \sigma=2\left (\frac{S_{d}}{Z}\right )^{(d-2)/d},\label{crit11}
\end{equation}
and where $Z$ is given by Eq. (\ref{sup6}). Proceeding along the
lines of \cite{virial1}, we could expand the solutions of
Eq. (\ref{crit6}) (or of the equivalent equation for $f$) in powers of
$\Theta^{-1}$ in the limit $\Theta\rightarrow +\infty$.

\begin{figure}[htbp]
\centerline{ \includegraphics[width=8cm,angle=0]{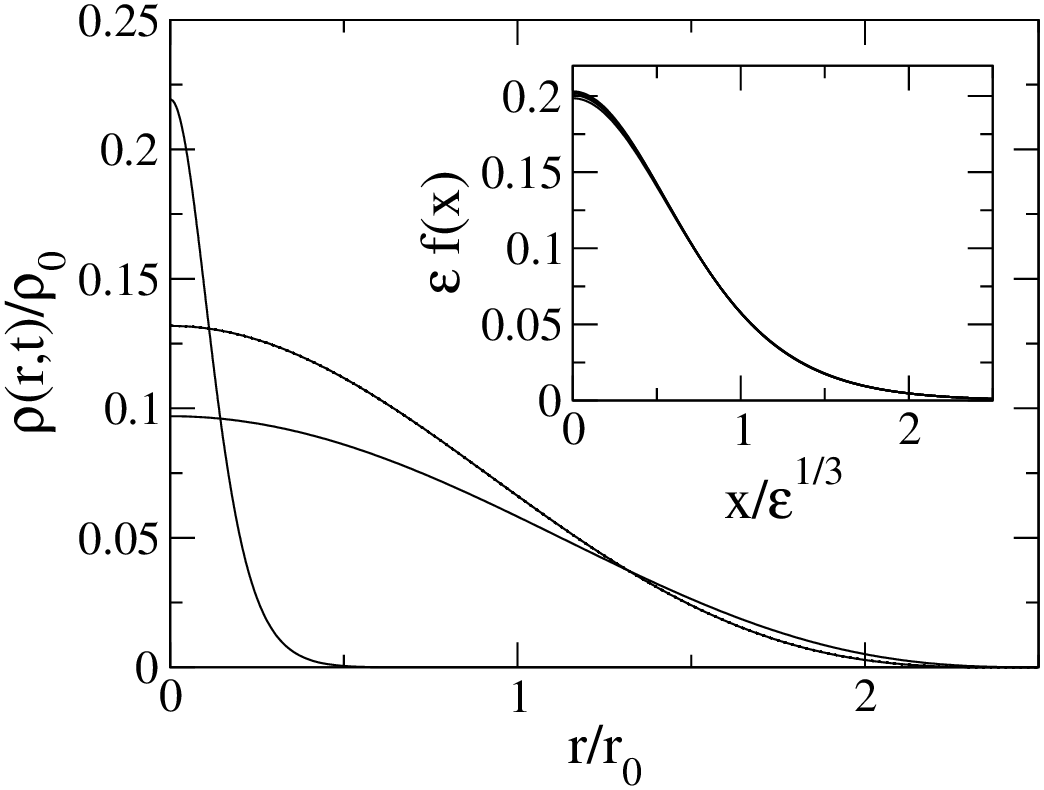} }
\caption[]{In $d=3$ and for $n=n_3=3$, we compare the scaling
profiles for $\Theta=0.21$ near $\Theta_c\approx 0.20872$,
$\Theta=1$, and $\Theta=100$ (top to bottom full lines; for clarity,
the $\Theta=0.21$ profile has been scaled down by a factor $150$).
For $\Theta\gg 1$, the invariant profile corresponds to the
Barenblatt solution (pure anomalous diffusion) which is a Tsallis
distribution with index $\gamma_{4/3}=1+1/n_3$. For
$\Theta\rightarrow \Theta_{c}$ the invariant profile tends to the
profile of a steady polytrope with index $n_3$.  For an intermediate
temperature $\Theta=1$, we illustrate the perfect observed data
collapse by plotting $r_{0}^{d}(t)\rho(r,t)$ as a function of
$r/r_0(t)$, for $t=1.5^n$ $(n=0,...,13)$. These 14 curves are
indistinguishable from the theoretical scaling profile. In the
insert, we illustrate the scaling relation of Eq.~(\ref{scathc})
obtained for different values of $\varepsilon=(\Theta-\Theta_c)/\Theta_c\to 0$.}
\label{evap2}
\end{figure}

In Fig.~\ref{evap2}, we show the form of the evaporation density
profile $f$ as a function of $\Theta>\Theta_c$. As $\Theta$
approaches $\Theta_c$, the central density diverges, whereas the
profile tends to the one corresponding to free diffusion for large
$\Theta$. In addition, we present numerical simulations for an
intermediate $\Theta$, showing that dynamical scaling is perfectly
obeyed. Moreover, when $\Theta\rightarrow \Theta_c$, we find that
the scaling function obeys itself a scaling relation (see insert of
Fig.~\ref{evap2}). Defining
$\varepsilon=(\Theta-\Theta_c)/\Theta_c$, we find
\begin{equation}
f(\Theta,x)=\varepsilon^{-1}F(x/\varepsilon^{1/d}),\label{scathc}
\end{equation}
where $F$ takes the form of a steady polytropic profile of index
$n_3$. This scaling relation implies that close to $\Theta_c$, the
$d$-th moment of $r$ scales as
\begin{equation}
\langle r^d(t)\rangle\sim (\Theta-\Theta_c)t,
\end{equation}
which is a generalization of our exact result for $d=2$
($n_3=+\infty$, $T_c=1/4$) \cite{virial1},
\begin{equation}
\langle r^2(t)\rangle= 4(T-T_c)t+\langle r^2(0)\rangle.
\end{equation}

\section{Analogy between the limiting mass of white dwarf stars
and the critical mass of bacterial populations}
\label{sec_analogy}

The generalized Smoluchowski-Poisson (GSP) system describing the
dynamics of self-gravitating Langevin particles shares many analogies
with the generalized Keller-Segel (GKS) model describing the
chemotaxis of bacterial populations. Below, we briefly review the
basic equations of chemotaxis and show the close link with the present
work.

\subsection{The generalized Keller-Segel model}
\label{sec_gks}

The original Keller-Segel model has the form \cite{ks}:
\begin{eqnarray}
\label{gks1}
\frac{\partial\rho}{\partial t}=\nabla\cdot
(D_{2}(\rho,c)\nabla\rho)-\nabla\cdot (D_{1}(\rho,c)\nabla c),
\end{eqnarray}
\begin{equation}
\label{gks2}\epsilon {\partial c\over\partial t}=-k(c) c+h(c) \rho+D_{c}\Delta c.
\end{equation}
The drift-diffusion equation (\ref{gks1}) governs the evolution of the
density of bacteria $\rho({\bf r},t)$ and the reaction-diffusion
equation (\ref{gks2}) governs the evolution of the secreted chemical
$c({\bf r},t)$. The bacteria diffuse with a diffusion coefficient
$D_{2}$ and they also move in a direction of a positive gradient of
the chemical (chemotactic drift). The coefficient $D_{1}$ is a measure
of the strength of the influence of the chemical gradient on the flow
of bacteria. On the other hand, the chemical is produced by the
bacteria with a rate $h(c)$ and is degraded with a rate $k(c)$. It
also diffuses with a diffusion coefficient $D_{c}$. In the primitive
Keller-Segel model, $D_{1}=D_{1}(\rho,c)$ and $D_{2}=D_{2}(\rho,c)$
can both depend on the concentration of the bacteria and of the
chemical. This can take into account microscopic constraints, like
close-packing effects \cite{ph,degrad,kin} or anomalous diffusion \cite{lang}.

If we assume a constant diffusion coefficient $D_2=D$ and a constant
mobility $D_1/\rho=\chi$ (we also consider a constant production
rate $\lambda$ and a constant degradation rate $k^2$ of the chemical),
we obtain the standard
Keller-Segel (KS) model
\begin{eqnarray}
\label{gks3}
\frac{\partial\rho}{\partial t}=\nabla\cdot \left
(D\nabla\rho-\chi \rho\nabla c \right ),
\end{eqnarray}
\begin{equation}
\label{gks4}\epsilon {\partial c\over\partial t}=\Delta c-k^2c+\lambda\rho.
\end{equation}
If we now assume that the diffusion coefficient  and the mobility
depend on the concentration of the bacteria, and if we set $D_2=D
h(\rho)$ and $D_1=\chi g(\rho)$, where $h$ and $g$ are positive
functions, we obtain the generalized Keller-Segel (GKS) model
\cite{ph,degrad,kin}:
\begin{eqnarray}
\label{gks5}
\frac{\partial\rho}{\partial t}=\nabla\cdot
(Dh(\rho)\nabla\rho-\chi g(\rho)\nabla c),
\end{eqnarray}
\begin{equation}
\label{gks6}\epsilon {\partial c\over\partial t}=\Delta c-k^2c+\lambda\rho.
\end{equation}
Equation (\ref{gks5}) can be viewed as a nonlinear mean field
Fokker-Planck (NFP) equation \cite{gfp} associated with a stochastic
process of the form
\begin{equation}
\frac{d{\bf r}}{dt}=\chi(\rho)\nabla c+\sqrt{2D(\rho)}{\bf R}(t),\label{gks7}
\end{equation}
with a diffusion coefficient
$D(\rho)=(D/\rho)\int^{\rho}h(\rho')d\rho'$ and a mobility
$\chi(\rho)=\chi g(\rho)/\rho$. These equations are associated with a
notion of effective generalized thermodynamics \cite{frank,gfp}. The Lyapunov
functional of the NFP equation (\ref{gks5})-(\ref{gks6}) can be
written in the form of a generalized free energy $F=E-T_{eff}S$ where
\begin{eqnarray}
\label{gks8}
E=\frac{1}{2\lambda}\int \left\lbrack (\nabla c)^{2}+k^{2} c^{2}\right \rbrack
\, d{\bf r}-\int \rho c \, d{\bf r},
\end{eqnarray}
is the energy, $T_{eff}=D/\chi$ is an effective temperature given by
an Einstein-like relation and
\begin{eqnarray}
\label{gsk9}
S=-\int C(\rho)\, d{\bf r}, \qquad C''(\rho)=\frac{h(\rho)}{g(\rho)},
\end{eqnarray}
is a generalized entropy. A straightforward calculation shows that
\begin{eqnarray}
\label{gks10}
\dot F=-\frac{1}{\lambda\epsilon}\int
(-\Delta c+k^{2}c-\lambda\rho)^{2} \,d{\bf r}\nonumber\\
-\int \frac{1}{\chi g(\rho)}(Dh(\rho)\nabla\rho-\chi g(\rho)\nabla c)^{2}\,d{\bf r}\le 0,
\end{eqnarray}
which is the expression of the $H$-theorem in the canonical ensemble
adapted to dissipative systems.  If we consider the particular case of
a constant mobility $g(\rho)=\rho$ and a power law diffusion
$h(\rho)=\gamma\rho^{\gamma-1}$, with $\gamma=1+1/n$, we obtain the
polytropic Keller-Segel model \cite{lang}:
\begin{eqnarray}
\label{gsk11}
\frac{\partial\rho}{\partial t}=\nabla\cdot \left
(D\nabla\rho^{\gamma}-\chi \rho\nabla c \right ),
\end{eqnarray}
\begin{equation}
\label{gks12}\epsilon {\partial c\over\partial t}=\Delta c-k^2c+\lambda\rho.
\end{equation}
The standard Keller-Segel model is recovered for $\gamma=1$.  Finally,
if we neglect the degradation of the chemical ($k=0$) and consider a
limit of large diffusivity of the chemical (implying $\epsilon=0$), we
obtain for sufficiently large concentrations (see
Appendix C of \cite{kin}):
\begin{eqnarray}
\label{gsk13}
\frac{\partial\rho}{\partial t}=\nabla\cdot \left
(D\nabla\rho^{\gamma}-\chi \rho\nabla c \right ),
\end{eqnarray}
\begin{equation}
\label{gsk14}\Delta c=-\lambda\rho.
\end{equation}
These equations are isomorphic to the generalized
Smoluchowski-Poisson (GSP) system (\ref{up1})-(\ref{up2}) provided
that we set
\begin{equation}
\label{gks15}D=K/\xi,\quad  \chi=1/\xi,\quad  c=-\Phi, \quad \lambda=S_{d}G.
\end{equation}
Therefore, the results of the present paper apply to the chemotactic
problem provided that the parameters are properly re-interpreted.

\subsection{Formulation of the results with the biological variables}
\label{sec_form}

In the gravitational context, we usually fix the coefficients $\xi$,
$G$ and $M$ and use the temperature $\Theta$ as a control parameter.
In the biological context, the coefficients $D$, $\chi$ and $\lambda$
are assumed given and the control parameter is the mass $M$.
Therefore, it may be useful to briefly reformulate the previous
results in terms of the mass, using notations adapted to the
chemotactic problem.

For the critical index $n=n_{3}=d/(d-2)$ in $d\ge 2$, the steady
states (polytropes) of the GKS model (\ref{gsk13})-(\ref{gsk14})
exist, in an unbounded domain, for a unique value of the mass given by \cite{csmasse}:
\begin{equation}
\label{f1}M_{c}=S_{d}\left \lbrack
\frac{D(1+n_3)}{\chi\lambda} \right\rbrack^{n_3/(n_3-1)}\omega_{n_3}.
\end{equation}
For $d=3$, we have
\begin{equation}
\label{f2}M_{c}=32\pi\omega_{3}\left (
\frac{D}{\chi\lambda} \right )^{3/2}\simeq 202.8956...\left (
\frac{D}{\chi\lambda} \right )^{3/2}.
\end{equation}
For $d=2$, using the identity (\ref{up18}), we recover the critical
mass
\begin{equation}
\label{f3}M_{c}=\frac{8\pi D}{\chi\lambda},
\end{equation}
associated with the two-dimensional standard Keller-Segel (KS) model
(see \cite{mt} and references therein). It is convenient to
introduce rescaled variables so that $D=\lambda=\chi=1$. With this
system of units the critical mass is $M_c(d)=S_d
(1+n_3)^{n_3/(n_3-1)}\omega_{n_3}=S_d\lbrack
2(d-1)/(d-2)\rbrack^{d/2}\omega_{d/(d-2)}$. For example, $M_c(d=2)=8\pi$ and
$M_c(d=3)=32\pi\omega_3=202.8956...$. Using the approximate
expression of $\omega_n$ obtained in Eq. (B72) of \cite{wd}, we can
derive an approximate expression of the critical mass in the form
\begin{equation}
\label{f4}M_{c}^{approx}(d)=\frac{S_d}{d}\lbrack d(d+2)\rbrack^{d/2}.
\end{equation}
For $d=2$, it returns the exact result
$M_{c}^{approx}(2)=M_c=8\pi$. On the other hand,
$M_c^{approx}(d=3)=243$ and $M_c^{approx}(d=4)=2842$. Using
$S_d=2\pi^{d/2}/\Gamma(d/2)$ we find that $M_c^{approx}(d)\sim
2\pi^{d/2}d^d/\Gamma(d/2)$ for $d\rightarrow +\infty$.

Let us briefly discuss the critical dynamics of the GKS system with
index $n=n_{3}=d/(d-2)$ for $d\ge 2$, depending on the total mass of the
bacteria.  For $M<M_c$, a box-confined system tends to an incomplete
polytrope confined by the walls of the box. In an unbounded domain,
the system evaporates in a self-similar way as discussed in
Sec. \ref{sec_crit}. For $M>M_c$, the system undergoes finite time
collapse as discussed in Sec. \ref{sec_collapse}. In a finite time
$t=t_{coll}$, it forms a Dirac peak containing a mass $M_c$ surrounded
by a collapsing halo evolving quasi self-similarly with a time-dependent
exponent $\alpha(t)$ tending extremely slowly to $\alpha=d$ as
$t\rightarrow t_{coll}$. Thus,
\begin{equation}
\label{f5}\rho({\bf r},t)\rightarrow M_{c}\delta({\bf r})+\chi({\bf r},t),
\end{equation}
where $\chi(r)$ behaves roughly as $r^{-d}$ for $r\rightarrow 0$. For
$M=M_c$, the situation is delicate and depends on the dimension of
space. For $d=2$, in a bounded domain, the steady state of the KS
model is a Dirac peak ($\rho_{0}=+\infty$). We have constructed in
\cite{sc} a self-similar solution tending to this Dirac peak in
infinite time. The central density increases exponentially rapidly. In
an infinite domain, the KS model admits an infinite family of steady
state solutions parameterized by their central density but the Dirac
peak ($\rho_{0}=+\infty$) is selected dynamically (the other solutions
have an infinite moment of inertia and, since the moment of inertia is
conserved when $M=M_{c}$, they cannot be reached from an initial
condition with a finite moment of inertia). We have constructed in
\cite{virial1} a self-similar solution tending to this Dirac peak in
infinite time (and ejecting a small amount of mass at large distances
so as to satisfy the moment of inertia constraint).  The central
density increases logarithmically rapidly. For $d>2$ and $M=M_c$, in a
bounded domain, the GKS model admits an infinite family of steady
state solutions parameterized by their central density or,
equivalently, by their natural radius $R_*$. We have found numerically
that the system tends to the polytrope where the density reaches zero
at the box radius ($R_*=R$).

Due to the analogy between gravity and chemotaxis \cite{crrs}, we find
that the critical mass of bacterial populations in the standard
Keller-Segel model in $d=2$ and in the generalized Keller-Segel model
in $d>2$ for the critical index $n=n_3$ shares some resemblance with
the Chandrasekhar mass of white dwarf stars. For example, the curves
of Figs. \ref{mrho} and \ref{alphaeta3} also represent the mass of the
bacterial aggregate as a function of the central density. As we have
seen, they are strikingly similar to the mass-central density relation
of white dwarf stars in Fig. \ref{massedensiteD3}. Therefore,
bacteria and white dwarf stars share deep analogies despite their very
different physical nature \cite{csmasse}.

\section{Conclusions and perspectives: the GSP system with a
relativistic equation of state}

In this paper, we have studied the critical dynamics, at the index
$n=n_3$, of the GSP system and GKS model describing self-gravitating
Langevin particles and bacterial populations. This study completes our
previous investigation \cite{lang} that was restricted to the cases
$n<n_3$ and $n>n_3$. We have seen that, at the index $n=n_3$, there
exists a critical mass $M_c$ (independent on the size of the system)
that is connected to the Chandrasekhar limiting mass of white dwarf
stars \cite{csmasse}. In order to strengthen this analogy, it would be
interesting to study the GSP system (\ref{gsp1})-(\ref{gsp2}) with the
equation of state (\ref{wdp2}) corresponding to relativistic white
dwarf stars. In fact, we can already describe qualitatively the
behavior of the solutions by using the results obtained here for
polytropes (see also the stability results obtained in \cite{wd} for
relativistic white dwarf stars).

For $d=1$ and $d=2$, there exists an equilibrium state (global minimum
of free energy) for all values of the mass $M$. Therefore, the GSP
system relaxes towards that steady state.

For $d=3$, there exists a critical mass
$M_{Chandra}=0.196701...({hc/G})^{3/2}/(\mu H)^{2}$. For
$M<M_{Chandra}$, the GSP system tends to a partially relativistic
white dwarf star (global minimum of free energy). For $M\ll
M_{Chandra}$, the density is small so that the equation of state
reduces to that of a polytrope of index $n=3/2$ (classical
limit). Therefore, the GSP system relaxes towards a classical white
dwarf star as described in Fig. 21 of \cite{lang}. For $M=M_{Chandra}$
the density becomes large so that the equation of state reduces to
that of a critical polytrope of index $n=3$ (ultra-relativistic
limit). We expect that the GSP system forms a Dirac peak of mass
$M_{Chandra}$ in infinite time. For $M>M_{Chandra}$, there is no
equilibrium state and the system collapses. When the density reaches
high values, the system becomes equivalent to a polytrope of index
$n=3$. Therefore, according to the present study, it forms in a finite
time a Dirac peak of mass $M_{Chandra}$ surrounded by a halo evolving
quasi self-similarly with an exponent $\alpha(t)$ converging very
slowly to $\alpha=3$.

For $d=4$, there exists a critical mass
$M_c=0.0143958...h^4/(m^2G^2\mu^3 H^3)$ discovered in \cite{wd}. For
$M<M_c$, the steady states are unstable and the system can either
collapse or evaporate (depending on the form of the initial density
profile and on the basin of attraction of the solution). In case of
evaporation, when the density reaches low values, the system becomes
equivalent to a polytrope of critical index $n_{3/2}=n_{3}=2$
(classical limit). In that case, it undergoes a self-similar
evaporation similar to that described in Sec. \ref{sec_crit} where
diffusion and gravity scale the same way. In case of collapse, when
the density reaches high values, the system becomes equivalent to a
polytrope of index $n_{3}'=4>n_3=2$ (ultra-relativistic limit). In
that case, it undergoes a self-similar collapse similar to that
described in \cite{lang}. For $M>M_c$, there is no steady state and
the system collapses in the way discussed previously (energy
considerations developed in \cite{wd} show that there is no
evaporation in that case).

For $d\ge 5$, there is no steady state and the system can either
collapse or evaporate.  In case of evaporation, when the density
reaches low values, the system becomes equivalent to a polytrope of
index $n_{3/2}>n_{3}$. In that case, it undergoes a self-similar
evaporation similar to that described in Sec. \ref{sec_sup} where
gravity becomes asymptotically negligible. In case of collapse, when
the density reaches high values, the system becomes equivalent to a
polytrope of index $n_{3}'>n_3$. In that case, it undergoes a
self-similar collapse similar to that described in \cite{lang}.

As we have already mentioned, the real dynamics of white dwarf stars
is not described by the GSP system, but is much more complicated.
However, we think that the study of this simple dynamical model is an
interesting first step before considering more complicated models. At
least, it reveals the great richness of the problem.  A next step
would be to take into account inertial effects and study the
(generalized) Kramers-Poisson system and the corresponding
hydrodynamic equations \cite{virial2}.

\appendix

\section{Virial theorem and free energy of critical polytropes}
\label{sec_virial}

The scalar Virial theorem for the GSP system reads \cite{virial1}:
\begin{equation}
\label{v1}
\frac{1}{2}\xi \frac{dI}{dt}=2E_{kin}+W_{ii},
\end{equation}
where $I=\int \rho r^{2}\,d{\bf r}$ is the moment of inertia,
$E_{kin}=(d/2)\int P d{\bf r}$ is the kinetic energy of the
microscopic motion and $W_{ii}=-\int \rho {\bf r}\cdot \nabla\Phi
\,d{\bf r}$ is the Virial. For $d=2$, $W_{ii}=-GM^2/2$ and for $d\neq
2$, $W_{ii}=(d-2)W$ where $W=(1/2)\int \rho\Phi d{\bf r}$ is the
potential energy. If the system is enclosed within a box, we must
add a term $-dP_{b}V$ on the right hand side, where $P_b$ is the
pressure against the box. In the following, we assume that the
system is unbounded so that $P_{b}=0$.

For a polytropic equation of state $P=K\rho^{\gamma}$, with
$\gamma=1+1/n$, the free energy (\ref{up4}) can be written
\begin{equation}
\label{v2}
F=\frac{2n}{d}E_{kin}+W-nKM.
\end{equation}
Therefore, the Virial theorem can be expressed in the form
\begin{equation}
\label{v3}
\frac{1}{2}\xi \frac{dI}{dt}=\frac{dF}{n}+W_{ii}-\frac{dW}{n}+dKM.
\end{equation}
For the critical index $n=n_{3}=d/(d-2)$, we get
\begin{equation}
\label{v4}
\frac{1}{2}\xi \frac{dI}{dt}=(d-2)F+W_{ii}-(d-2)W+dKM.
\end{equation}
For $d\neq 2$, it reduces to
\begin{equation}
\label{v5}
\frac{1}{2}\xi \frac{dI}{dt}=(d-2)F+dKM.
\end{equation}
For a steady state ($\dot I=0$),  the Virial theorem implies
\begin{equation}
\label{v6}
F_{eq}=-\frac{d}{d-2}KM.
\end{equation}
We have seen in Sec. \ref{sec_up} that spherically symmetric steady
states of the GSP system with $n=n_3$ exist for a unique value of the
mass $M=M_{c}$ (for fixed $K$) or a unique value of the temperature
$\Theta=\Theta_{c}$ (for fixed $M$) and form an infinite family of
solutions parameterized by their central density $\rho_0$. According to
Eq. (\ref{v6}), they all have the same free energy, independent on the
central density $\rho_0$. Therefore, thermodynamical arguments do not
allow to select a particular solution among the whole family.

For $d=2$, the critical index $n_3\rightarrow +\infty$ and the
equation of state is isothermal with $K=k_BT/m$. Then, the Virial
theorem (\ref{v4}) becomes \cite{virial1}:
\begin{equation}
\label{v7}
\frac{1}{2}\xi \frac{dI}{dt}=2Nk_{B}(T-T_c),
\end{equation}
with $k_BT_c=GMm/4$. For a steady state ($\dot I=0$), the Virial
theorem implies $T=T_c$ or $M=M_c$. It directly yields the result
that unbounded two-dimensional isothermal spheres exist for a unique value of
the mass or temperature. The spherically symmetric solution is given
by Eq. (\ref{i14}) reading
\begin{equation}
\label{v8}
\rho(r)=\frac{\rho_{0}}{\lbrack 1+(\pi\rho_{0}/M)r^2\rbrack^2}.
\end{equation}
This family of steady solutions is parameterized by the central density
$\rho_{0}$.  The corresponding mass profile is given by
$M(r)=\int_{0}^{r}\rho(r')2\pi r' dr$ and the gravitational potential
can be obtained from the Gauss theorem $d\Phi/dr=GM(r)/r$ with the
gauge condition $\Phi(r)\sim GM\ln r$ for $r\rightarrow +\infty$. This
yields
\begin{equation}
\label{v9}
M(r)=\frac{\pi \rho_{0} r^2}{1+(\pi\rho_{0}/M)r^2},
\end{equation}
\begin{equation}
\label{v9b}
 \Phi(r)=\frac{GM}{2}\ln\left (\frac{M}{\pi\rho_{0}}+r^2\right ).
\end{equation}
From these expressions, we find that the potential energy is
\begin{equation}
\label{v10}
W=\frac{GM^2}{4}\left\lbrack 1+\ln \left (\frac{M}{\pi \rho_{0}}\right )\right\rbrack.
\end{equation}
On the other hand, the Boltzmann entropy $S_{B}=-k_{B}\int
(\rho/m)\ln(\rho/m) d{\bf r}$ can be written
\begin{equation}
\label{v11}
S_{B}=2Nk_B\left\lbrack 1-\frac{1}{2}\ln \left (\frac{\rho_0}{m}\right )\right \rbrack.
\end{equation}
Therefore, the Boltzmann free energy $F_{B}=W-TS_{B}$ is given by
\begin{equation}
\label{v12}
F_{B}=-\frac{GM^2}{4}\left\lbrack 1+\ln \left (\frac{\pi}{N}\right )\right\rbrack.
\end{equation}
We conclude that the free energy of unbounded  isothermal spheres in
two dimensions is independent on the central density $\rho_{0}$.

\section{An equation for the moments $\langle r^k\rangle$}
\label{m}

Let us introduce the moments of order $k$:
\begin{equation}
\label{m1}
I_{k}(t)=\int\rho r^k d{\bf r}.
\end{equation}
For $k=2$, we recover the moment of inertia.  Taking the time
derivative of Eq. (\ref{m1}), using the generalized Smoluchowski
equation (\ref{gsp1}) and integrating by parts, we obtain
\begin{equation}
\label{m2} \frac{1}{k}\xi\frac{dI_{k}}{dt}=-\int r^{k-2}{\bf r}\cdot
\nabla P \, d{\bf r}-\int r^{k-2}\rho \ {\bf r}\cdot \nabla\Phi \,
d{\bf r}.
\end{equation}
Integrating by parts the first term, we get
\begin{equation}
\label{m3} \frac{1}{k}\xi\frac{dI_{k}}{dt}=(k+d-2)\int  P r^{k-2} \,
d{\bf r}-\int r^{k-2}\rho {\bf r}\cdot \nabla\Phi \, d{\bf r}.
\end{equation}
If we take into account boundary effects, we need to introduce a
pressure term $-\oint Pr^{k-2}{\bf r}\cdot d{\bf S}$ on the r.h.s.
For $k=2$, we recover the Virial theorem (\ref{v1}). On the other
hand, for a spherically symmetric system, using the Gauss theorem, the
second integral can be simplified and we obtain
\begin{equation}
\label{m4} \frac{1}{k}\xi\frac{dI_{k}}{dt}=(k+d-2)\int  P r^{k-2}
d{\bf r}-G \int_{0}^{+\infty} r^{k-d}M(r)\frac{\partial M}{\partial
r}dr.
\end{equation}
For $k=d$, the second integral can be calculated explicitly and we get
\begin{equation}
\label{m5}
\frac{1}{d}\xi\frac{dI_{d}}{dt}=2(d-1)\int  P r^{d-2} d{\bf r}-\frac{GM^2}{2}.
\end{equation}
For $d=1$, the first term on the r.h.s. must be replaced by $2P(0,t)$.


\begin{thebibliography}{}

\bibitem{antonov} {\small V.A. Antonov, Vest. Leningr. Gos. Univ. {\bf 7}, 135 (1962).}

\bibitem{lb} {\small D. Lynden-Bell, Extrait du Bulletin Astronomique {\bf 3}, 305 (1968); D. Lynden-Bell \& R. Wood, Mon. Not. R. Astron. Soc. {\bf 138}, 495 (1968).}

\bibitem{thirring} {\small W. Thirring, Z. Phys. {\bf 235}, 339 (1970).}

\bibitem{paddy} {\small T. Padmanabhan, Phys. Rep.  {\bf 188}, 285
(1990).}

\bibitem{ijmpb} {\small P.H. Chavanis, Int. J. Mod. Phys. B, {\bf
20}, 3113 (2006).}

\bibitem{houches} {\small {\it Dynamics and thermodynamics of systems
with long range interactions}, edited by T. Dauxois {\it et al.}, Lecture Notes in Physics {\bf 602}, Springer
(2002).}

\bibitem{assise} {\small {\it Dynamics and thermodynamics of systems
with long range interactions: Theory and experiments}, edited by
A. Campa {\it et al.}, AIP
Conf. Proc. {\bf 970} (AIP, 2008). }

\bibitem{artemiev} {\small A.I. Artemiev, I.E. Mazets, G. Kurizki, D. O'Dell, Int. J. Mod. Phys. B, {\bf
18}, 2027 (2004).}

\bibitem{ht} {\small P. Hertel, W. Thirring, in {\it Quanten und Felder}, ed. H.P. D\"urr (Vieweg, Braunschweig, 1971).}

\bibitem{pt} {\small P.H. Chavanis, Phys. Rev. E {\bf 65}, 056123 (2002).}

\bibitem{crs} {\small P.H. Chavanis, C. Rosier, C. Sire,
Phys. Rev. E {\bf 66}, 036105 (2002).}

\bibitem{sc} {\small C. Sire, P.H. Chavanis, Phys. Rev. E {\bf 66},
046133 (2002).}

\bibitem{post} {\small C. Sire, P.H. Chavanis, Phys. Rev. E {\bf 69},
066109 (2004).}

\bibitem{tcoll} {\small P.H. Chavanis, C. Sire,  Phys. Rev. E {\bf 70},
026115 (2004).}

\bibitem{multi} {\small J. Sopik, C. Sire, P.H. Chavanis,
Phys. Rev. E {\bf 72}, 026105 (2005).}


\bibitem{virial1} {\small P.H. Chavanis, C. Sire, Phys. Rev. E {\bf
73}, 066103 (2006). }

\bibitem{virial2} {\small P.H. Chavanis, C. Sire, Phys. Rev. E {\bf
73}, 066104 (2006). }

\bibitem{emden} {\small R. Emden, {\it Gaskugeln} (Teubner Verlag, Leipzig, 1907).}

\bibitem{chandrab} {\small S. Chandrasekhar, {\it An Introduction to the Theory of Stellar Structure} (Dover, New York, 1942).}

\bibitem{ks} {\small E. Keller, L.A. Segel J. theor. Biol. {\bf 26},
399 (1970).}

\bibitem{jl} {\small W. J\"ager, S. Luckhaus,
Trans. Am. Math. Soc. {\bf 329}, 819 (1992).}

\bibitem{murray} {\small J.D. Murray, {\it Mathematical Biology}
(Springer, Berlin, 1991).}

\bibitem{acedo}  {\small L. Acedo, Europhysics Letters  {\bf 73}, 698 (2006).}

\bibitem{nanjundiah}  {\small V. Nanjundiah, J. Theoret. Biol.   {\bf 42}, 63 (1973).}


\bibitem{cp} {\small S. Childress, J.K. Percus, Math. Biosci. {\bf
56}, 217 (1981).}

\bibitem{childress} {\small S. Childress, Lecture Notes in
Biomath. {\bf 55}, 61 (1984).}

\bibitem{nagai} {\small T. Nagai, Adv. Math. Sci. Appl.  {\bf 5}, 581
(1995).}

\bibitem{biler95} {\small P. Biler, Studia Mathematica   {\bf 114}, 181
(1995).}

\bibitem{herrero96} {\small M.A. Herrero, J.J.L. Velazquez, Math. Ann.
{\bf 306}, 583 (1996).}

\bibitem{herrerobio} {\small M.A. Herrero, J.J.L. Velazquez, J. Math. Biol.
{\bf 35}, 177 (1996).}

\bibitem{othmer}  {\small H. G. Othmer and A. Stevens, SIAM J. Appl. Math.    {\bf 57}, 1044 (1997).}

\bibitem{herrero97}  {\small M.A. Herrero, E. Medina and J.L. Velazquez, Nonlinearity  {\bf 10}, 1739 (1997).}

\bibitem{herrero98}  {\small M.A. Herrero, E. Medina, and J.L. Velazquez, J. Comput. Appl. Math.  {\bf 97}, 99 (1998).}

\bibitem{biler}  {\small P. Biler, Adv. Math. Sci. Appl.    {\bf 8}, 715 (1998).}

\bibitem{brenner}  {\small M.P. Brenner, P. Constantin, L.P. Kadanoff, A. Schenkel and S.C. Venkataramani, Nonlinearity  {\bf 12}, 1071 (1999).}

\bibitem{nagai2}  {\small T. Nagai, J. Inequal. Appl.   {\bf 6}, 37 (2001).}

\bibitem{rosier}  {\small C. Rosier, C.R. Acad. Sci. Paris S\'erie I   {\bf 332}, 903 (2001).}

\bibitem{bn}  {\small P. Biler, T. Nadzieja, Rep. Math. Phys.   {\bf 52}, 205 (2003).}

\bibitem{horstmann}  {\small D. Horstmann, Jahresberichte der DMV  {\bf 106}, 51 (2004).}

\bibitem{dolbeault} {\small J. Dolbeault, B. Perthame,
C. R. Acad. Sci. Paris, Ser. I {\bf 339}, 611 (2004).}

\bibitem{biler4}  {\small P. Biler, M. Cannone, I.A. Guerra, G. Karch,  Math. Ann.   {\bf 330}, 693 (2004).}

\bibitem{corrias} {\small L. Corrias, B. Perthame, H. Zaag,
Milan J. Math. {\bf 72}, 1 (2004).}

\bibitem{biler1} {\small P. Biler, G. Karch, P. Lauren\c{c}ot,
T. Nadzieja, Topol. Methods Nonlinear Anal. {\bf 27}, 133 (2006).}

\bibitem{biler2} {\small P. Biler, G. Karch, P. Lauren\c{c}ot,
T. Nadzieja, Math. Methods Appl. Sci.  {\bf 29}, 1563 (2006).}

\bibitem{blanchet1} {\small A. Blanchet, J.A. Carrillo, N. Masmoudi, to appear in Comm. Pure Appl. Math.}

\bibitem{blanchet2} {\small A. Blanchet, J. Dolbeault, B. Perthame, Electron. J. Differential Equations {\bf 44}, 32 (2006).}

\bibitem{souplet} {\small N. Kavallaris, P. Souplet, [arXiv:0804.4549].}

\bibitem{gen}  {\small P.H. Chavanis, Phys. Rev. E {\bf 68}, 036108 (2003).}

\bibitem{frank} {\small T.D. Frank, {\it Non linear Fokker-Planck equations} (Springer, Berlin, 2005).}

\bibitem{gfp}  {\small P.H. Chavanis, Eur. Phys. J. B {\bf 62}, 179 (2008).}

\bibitem{lang} {\small P.H. Chavanis, C. Sire, Phys. Rev. E {\bf
69}, 016116 (2004).}

\bibitem{logotropes} {\small P.H. Chavanis, C. Sire, Physica A {\bf
375}, 140 (2007).}

\bibitem{crrs} {\small P.H. Chavanis, M. Ribot, C. Rosier, C. Sire, Banach Center Publ. {\bf 66}, 103 (2004).}

\bibitem{bln} {\small P. Biler, P. Lauren\c{c}ot, T. Nadzieja, Adv. Differential Equations {\bf 9}, 563 (2004).}

\bibitem{degrad} {\small P.H. Chavanis, Eur. Phys. J. B {\bf 54}, 525
(2006).}

\bibitem{tsallis}  {\small C. Tsallis, J. Stat. Phys. {\bf 52}, 479 (1988).}

\bibitem{fowler}  {\small R.H. Fowler, Mon. Not. R. Astron. Soc. {\bf 87}, 114 (1926).}

\bibitem{chandra} {\small S. Chandrasekhar, Astrophys. J.
{\bf 74}, 81 (1931).}

\bibitem{csmasse} {\small P.H. Chavanis, C. Sire,  Physica A {\bf 387}, 1999 (2008). }

\bibitem{mt} {\small P.H. Chavanis, Physica A {\bf
384}, 392 (2007).}

\bibitem{wd} {\small P.H. Chavanis, Phys. Rev. D {\bf 76}, 023004
(2007).}

\bibitem{bcl} {\small A. Blanchet, J. A. Carrillo, P. Lauren\c{c}ot, submitted.}

\bibitem{chandraM} {\small S. Chandrasekhar, Mon. Not. R. Astron. Soc.
{\bf 95}, 207 (1935).}

\bibitem{lemou} {\small P.H. Chavanis, P. Lauren\c cot, M. Lemou, Physica A
{\bf 341}, 145 (2004).}

\bibitem{aaantonov} {\small P.H. Chavanis, A\&A
{\bf 451}, 109 (2006).}

\bibitem{camm} {\small G.L. Camm,  Mon. Not. R. Astron. Soc.
{\bf 110}, 305 (1950).}

\bibitem{katz} {\small J. Katz, Found. Phys. {\bf 33}, 223 (2003).}

\bibitem{ostriker} {\small J. Ostriker,  ApJ
{\bf 140}, 10560 (1964).}

\bibitem{borland} {\small L. Borland,  Phys. Rev. E
{\bf 57}, 6634 (1998).}

\bibitem{barenblatt} {\small  G. I. Barenblatt, V. M. Entov, and V. M. Ryzhik,
{\it Theory of Fluid Flows through Natural Rocks} (Kluwer Academic, Dordrecht, 1990).}

\bibitem{pp} {\small A.R. Plastino, A. Plastino, Physica A {\bf 222}, 347  (1995).}

\bibitem{ph}  {\small T. Hillen, K. Painter, Adv. Appl. Math.  {\bf 26}, 280  (2001).}

\bibitem{kin} {\small P.H. Chavanis, C. Sire, Physica A {\bf 384}, 199  (2007).}

\end{thebibliography}
\end{document}